\RequirePackage{fix-cm}
\documentclass[twocolumn,epjc3]{svjour3}  
\smartqed  
\RequirePackage{graphicx}
\usepackage{graphicx}
\usepackage{subcaption}
\usepackage{xcolor}

\usepackage[english]{babel}
\RequirePackage{latexsym}
\usepackage{amsmath}
\usepackage{amssymb}
\RequirePackage[numbers,sort&compress]{natbib}
\usepackage[
  colorlinks=true,
  linkcolor=blue,
  citecolor=magenta,
  urlcolor=blue
]{hyperref}
\usepackage{hyphenat}
\usepackage[displaymath, mathlines]{lineno}

\journalname{Eur. Phys. J. }

\begin{document}
\title{A rapid low-background assay of $^{210}$Pb in archaeological lead}
\author{
M.~Consonni\thanksref{1, 2,a}
\and
M.~Clemenza\thanksref{1, 2, b}
\and
E.~Di~Stefano\thanksref{2, 3}
\and
N.~Ferreiro Iachellini\thanksref{1, 2}
\and
F.~Filippini\thanksref{1,2,3}
\and
A.~Gardini\thanksref{LENA, INFN-PV}
\and
G.~Grosso\thanksref{LENA}
\and
L.~Pattavina\thanksref{1, 2}
\and
R.~Della~Pergola\thanksref{3}
\and
S.~Quitadamo\thanksref{1, 2}
\and
E.~Sala\thanksref{2, 4}
\and
F.~Saliu\thanksref{2, 3}
\and
A.~Salvini\thanksref{LENA, INFN-PV}
\and
L.~Trombetta\thanksref{1, 2}
}
\thankstext{a}{e-mail: m.consonni29@campus.unimib.it}
\thankstext{b}{e-mail: massimiliano.clemenza@mib.infn.it}



\institute{Dipartimento di Fisica, Universit\`a di Milano - Bicocca, Piazza della Scienza 3, I-20126 Milano, Italy \label{1}
\and
INFN Sezione di Milano - Bicocca, Piazza della Scienza 3, I-20126 Milano, Italy \label{2}
\and
DISAT, Universit\`a di Milano - Bicocca, Piazza della Scienza 1, I-20126 Milano, Italy \label{3}
\and
Laboratorio Energia Nucleare Applicata, Via Aselli 41, I-27100 Pavia, Italy \label{LENA}
\and
INFN Sezione di Pavia, Via Bassi 6,  I-27100 Pavia, Italy \label{INFN-PV}
\and
Center for Underground Physics, Institute for Basic Science, 34126 Daejeon, Korea \label{4}
}
\date{Received: date / Revised version: date}
%

\maketitle

\abstract{
In this work, we present a fast and highly efficient method for the measurement of $^{210}$Pb in metallic archaeological lead using the commercial low--background liquid scintillation counter Wallac Quantulus~1220 installed at the University of Milano-Bicocca (Italy). By combining an optimized chemical preparation with pulse--shape analysis (PSA), the technique achieves sensitivities at the level of a few $10^2$~mBq/kg within one week of measurement, using sample masses below 1~g.
The method enables the simultaneous identification of the $\beta$ decays of $^{210}$Pb and $^{210}$Bi and the $\alpha$ decay of $^{210}$Po, allowing a direct verification of secular equilibrium within the decay chain. With extended acquisition times, detection limits below 100~mBq/kg are reached after approximately 40~days.

This approach provides a rapid, accessible, and reliable tool for the radiopurity screening of lead, and is well suited for quality control and R\&D activities in next--generation low--background and rare--event physics experiments. Moreover, the method has the potential to be extended to other materials relevant for low--background applications.

} 
\section{Introduction}

The realization of ultra--low--background experiments is of paramount importance in rare--event physics, where processes such as neutrinoless double--beta decay ($0\nu\beta\beta$)~\cite{neutrinoless}, dark--matter--nucleus interactions~\cite{darkmatter}, and neutrino--nucleus scattering~\cite{neutrinos} can be completely overwhelmed by background events generated by environmental radioactivity and cosmic rays.

Natural radioactivity is unavoidably present in all materials used in detector construction, making it challenging to achieve the ultra--low--background conditions required to detect such elusive signals. Moreover, the massive shielding employed to reduce the environmental radioactivity background may itself become the dominant background source if it is not made of radiopure materials, thereby severely limiting detector sensitivity. For this reason, material assay is a critical aspect of low counting rate experiments~\cite{Laubenstein:2020rbe}.

The primary background--mitigation strategy adopted by ultra--low--background experiments is the installation of detectors deep underground, in order to suppress the flux of cosmic rays. The surrounding rock provides nearly complete solid--angle shielding against cosmic radiation; nevertheless, detectors remain sensitive to environmental radioactivity, making additional local shielding necessary.

Large--volume detectors such as JUNO~\cite{juno2022juno}, KamLAND--Zen~\cite{Kamland2022}, XENONnT~\cite{xenon2024xenonnt}, and DarkSide-20k~\cite{zani2024darkside} employ highly purified liquids, which are readily available in large quantities and can act both as passive shielding and as active veto detectors.

In contrast, small--volume detectors operating at cryogenic temperatures, such as CUORE~\cite{cuore2018first}, AMORE~\cite{Amore-2024}, and CRESST~\cite{CRESSTIII2019first}, cannot accommodate large liquid shields due to technical constraints imposed by the cryogenic infrastructure. These experiments are therefore designed to be compact and to operate with limited shielding volumes. In this context, lead (Pb) is commonly used as the primary shielding material and is placed in close proximity to the detector active volume.

Lead has also been employed as a target material in neutrino experiments. The OPERA experiment~\cite{OPERA2009detection} measured $\nu_\mu\!-\!\nu_\tau$ oscillations using Pb plates interleaved with nuclear emulsions as targets for the CNGS muon--neutrino beam. Pb is currently used in the HALO experiment~\cite{Vaananen:2011bf}, which aims to detect neutrinos from supernovae (SN) using a Pb target coupled to $^3$He neutron counters. Pb will also be employed in the upcoming RES--NOVA experiment~\cite{RESNOVA_Pb,Pattavina_2021}, a cryogenic detector designed to measure neutrinos from astrophysical sources (e.g.\ SN and the Sun) as well as dark matter (DM) candidates~\cite{alloni2025new}, using PbWO$_4$ crystals produced from archaeological Pb~\cite{kg-scale}.

For these reasons, the search for highly radiopure Pb has been a long--standing topic of interest in rare--event physics, particularly when Pb is used as an active detector component. The dominant radioactive contaminant in Pb is $^{210}$Pb, a daughter of the $^{238}$U decay chain and a ubiquitous environmental radionuclide. Since chemical purification techniques are ineffective at removing $^{210}$Pb from metallic Pb, the only viable approach to reduce its activity is radioactive decay, which requires timescales of several hundreds of years ($\tau_{1/2}(^{210}\mathrm{Pb}) = 22.3$~y). Consequently, archaeological Pb, such as material recovered from sunken ships, where cosmogenic activation is strongly suppressed, represents a valuable source of highly radiopure Pb~\cite{pattavina2019radiopurity}.

In this work, we present a rapid and highly efficient method for the measurement of $^{210}$Pb in metallic Pb using a commercial low--background liquid scintillation counter, the Wallac Quantulus~1220 manufactured by PerkinElmer. The experimental setup, combined with a dedicated sample preparation procedure, enables sensitivities at the level of 100~mBq/kg with exposures of only a few $\mathrm{g}\cdot \mathrm{d}$.

This technique is well suited for the rapid assessment of Pb radiopurity before and after chemical purification, representing a valuable tool for research and development activities in rare--event physics experiments employing Pb as shielding or target material. Furthermore, the method allows the simultaneous detection of the two $\beta$ decays of $^{210}$Pb and $^{210}$Bi, as well as the $\alpha$ decay of $^{210}$Po, enabling verification of secular equilibrium among the three radionuclides in the decay chain (see Fig.~\ref{210Pb_scheme}).

\section{Archaeological Pb as background source}

Archaeological Pb is a particularly attractive raw material due to its expected high radiopurity \cite{alessandrello1991measurements, Laubenstein:2020rbe, kg-scale}, especially with respect to $^{238}$U and $^{232}$Th \cite{alessandrello1991measurements, Boiko} and their decay products. This is generally attributed both to the ancient refining processes performed by historical producers (Lead in the Roman period was extensively produced and refined, often as a by-product of silver extraction, as documented by archaeometallurgical and isotopic studies \cite{bode2021roman, bode2009tracing}) and the long elapsed time since production. Samples recovered from shipwrecks are even more valuable, as they benefit from the prolonged shielding provided by seawater overburden, which significantly reduces the flux of cosmic-ray particles, in particular the hadronic component responsible for cosmogenic activation \cite{Laubenstein:2020rbe}. Nevertheless, the use of archaeological Pb in rare--event physics may still be limited by the possible presence of $^{210}$Pb \cite{Laubenstein:2020rbe, pattavina2019radiopurity}.

Whenever $^{210}$Pb is embedded in a detector, it induces different types of background signatures (see Fig.~\ref{210Pb_scheme}):

\begin{figure}
    \centering
    \includegraphics[width=0.8\linewidth]{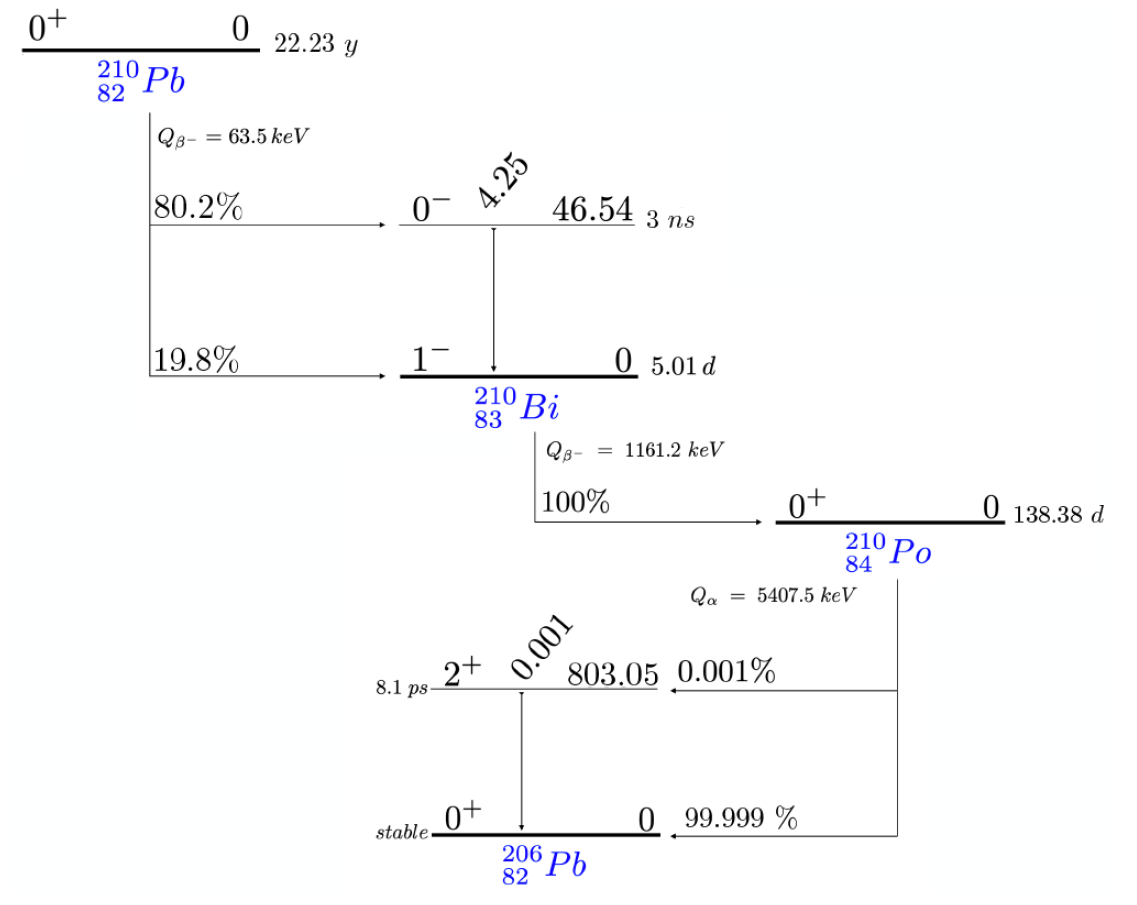}
    \caption{Decay scheme of the $^{210}${Pb} decay chain. For the $^{210}${Pb} and $^{210}${Bi} only the $\beta^-$ branches are shown, since $\alpha$ decays have far smaller branching ratios.}
    \label{210Pb_scheme}
\end{figure}

\begin{itemize}
\item $^{210}$Pb decays via $\beta^-$ emission with an endpoint energy of 63~keV \cite{lnhb_lara}. This decay channel represents one of the most relevant background sources for experiments operating at low energy thresholds, such as those searching for DM or aiming at the observation of SN neutrinos via nuclear recoils \cite{darkmatter, neutrinos}. The $\beta^-$ decay can populate an excited state of $^{210}$Bi at 46~keV, which subsequently de-excites by emitting a 46~keV $\gamma$ ray before reaching the ground state.

\item $^{210}$Bi decays through a $\beta^-$ channel with an endpoint energy of 1.16~MeV. The emitted energetic electron can produce bremsstrahlung radiation as well as X-rays with energies up to about 90~keV while traversing the detector volume \cite{orrell2016assay, knoll2010radiation}.

\item $^{210}$Po decays by emitting a 5.3~MeV $\alpha$ particle, which constitutes a relevant background for neutrinoless double--beta decay ($0\nu\beta\beta$) experiments~\cite{Clemenza:2011zz, Laubenstein:2020rbe}. This decay also represents a direct background source for experiments searching for DM. In addition, due to their short range, $\alpha$ particles can induce ($\alpha$,n) reactions in low--$Z$ elements present in contaminated materials, leading to the production of neutrons. Neutrons are a particularly dangerous background, as they deposit energy through nuclear recoils, thereby mimicking the expected DM signal. Furthermore, neutrons produced via ($\alpha$,n) reactions can activate surrounding materials, resulting in an additional increase of the background.
\end{itemize}

Due to its relatively short half--life (22.3~y), $^{210}$Pb is typically measured using radiometric techniques \cite{Laubenstein:2020rbe}:

\begin{itemize}
\item Direct $\gamma$ counting of the 46~keV photon emitted in 4.25\% of $^{210}$Pb decays \cite{heusser2006low, Laubenstein:2020rbe}.  
This technique suffers from several limitations, primarily the low branching ratio of the $\gamma$ transition and its low energy, which enhances self--absorption in the sample and results in a reduced detection efficiency for high--purity germanium (HPGe) detectors. Therefore, it is necessary to use kg--scale samples and custom--made facilities able to accommodate large, heavy samples to achieve sensitivities of $\sim$ Bq/kg\cite{alessandrello1991measurements, heusser2006low}.

\item $\beta$ counting of $^{210}$Bi ($t_{1/2} = 5$~d),  the immediate daughter of $^{210}$Pb, by Cherenkov counting of the $^{210}$Bi $\beta$ in liquid matrices \cite{cuesta2022comparative, stojkovic2020210pb}, or by detection of bremsstrahlung $\gamma$ using HPGe~\cite{orrell2016assay}.  
These methods require extensive chemical preparation to separate $^{210}$Bi from other radionuclides, as many of them contribute to the continuous $\beta$ spectrum and may act as background sources.

As for bremsstrahlung $\gamma$ detection, accuracy of these measurements relies upon $^{210}$Po $\alpha$--spectroscopy \cite{orrell2016assay}.
To infer the $^{210}$Pb activity from $^{210}$Bi $\beta$ counting, secular equilibrium between the two nuclides must be established. Owing to the much shorter half--life of $^{210}$Bi, this equilibrium is reached on relatively short timescales.

\item $\alpha$ counting of the $^{210}$Po daughter ($t_{1/2} = 138$~d) \cite{Quantulus_sea2016rapid, Quantulus_geo2004simple}.  
The main drawback of this technique is that secular equilibrium cannot always be assumed, as chemical processing may introduce different amounts of $^{210}$Pb and $^{210}$Po due to their different chemical yields \cite{danon1956solvent}.  
Nevertheless, $\alpha$ detection is an effective method for geological or archaeological samples, for which secular equilibrium can generally be assumed to hold \cite{Quantulus_geo2004simple}.

$^{210}$Po $\alpha$ counting can also be performed using bolometers \cite{alessandrello1993measurements, alessandrello1998measurements, pattavina2019radiopurity}. This technique provides the highest detection sensitivity, however it requires costly infrastructures and can be applied only to metallic lead samples.
\end{itemize}

The radiopurity of ancient Pb samples has been extensively investigated \cite{alessandrello1991measurements, alessandrello1993measurements, alessandrello1998measurements, orrell2016assay, heusser2006low, pattavina2019radiopurity}, revealing very low levels of intrinsic $^{210}$Pb contamination. However, a low $^{210}$Pb activity alone is not a sufficient condition to qualify Pb as a candidate material for rare--event physics \cite{Laubenstein:2020rbe}. In fact, the presence of $^{238}$U and $^{232}$Th can feed the lower part of their respective decay chains, where Pb radioisotopes are present. Nevertheless, several studies have shown that, in archaeological Pb, the concentrations of $^{238}$U and $^{232}$Th are lower than those observed in low--activity modern Pb~\cite{alessandrello1991measurements}. For these reasons, archaeological Pb represents a particularly valuable material for rare--event physics experiments.

The most stringent limit on the $^{210}$Pb concentration in ancient Pb to date was reported in Ref.~\cite{pattavina2019radiopurity}, reaching values below $715~\mu\mathrm{Bq/kg}$ at 90\% C.L. This result was obtained using a 10~g ancient Pb sample operated as the absorber of a cryogenic detector, by measuring the $\alpha$ decays of $^{210}$Po.  
A summary of the most relevant measurements performed on archaeological Pb samples is reported in Table~\ref{210Pb_table}.

\begin{table}[t]
\centering
\caption{Summary of the most sensitive measurements of $^{210}$Pb concentrations in ancient and low--background Pb samples. Limits are reported at 90\% C.L..}
\label{210Pb_table}
\begin{tabular}{lcccc}
\hline
Detector & $^{210}$Pb & Mass & Measurement time & Ref. \\
 & [mBq/kg] & [kg] & [d] & \\
\hline
HPGe            & $< 900$  & 4.5   & 9.83  & \cite{alessandrello1991measurements} \\
HPGe            & $< 1300$ & 22.1  & 37    & \cite{heusser2006low} \\
Planar Si       & 100      & 0.01  & 14    & \cite{orrell2016assay} \\
Bolometer       & $< 20$   & 0.01  & n/a   & \cite{alessandrello1993measurements} \\
Bolometer       & $< 4$    & 0.01  & 5     & \cite{alessandrello1998measurements} \\
Bolometer       & $< 0.7$  & 0.01  & 12.5  & \cite{pattavina2019radiopurity} \\
\hline
\end{tabular}
\end{table}

\section{Archaeological Pb processing}

Archaeological Pb is typically recovered in the form of ingots or bricks. As a result, a sequence of purification and refining steps is required to convert archaeological Pb into forms and purity levels suitable for technical applications, such as detector shielding or crystal production~\cite{PMO_pattavina,PMO_Korea,PWO,kg-scale,FerreiroIachellini:2021qgu}.

Several purification and refining techniques can be applied to Pb ingots, including recasting, Pb atomization, and zone refining, with the aim of removing chemical impurities from the metal~\cite{Boiko}. These techniques exploit differences in melting points and mobilities of the various impurities and preserve the archaeological Pb in metallic form, either as ingots or powders.

However, during any purification process there is a risk of contamination of archaeological Pb, in particular due to the presence of $^{222}$Rn ($t_{1/2} = 3.8$~d) in air, which rapidly decays into $^{210}$Pb. Such contaminations can be redistributed and potentially concentrated by refining techniques, effectively becoming bulk contaminations. In addition, any pre-existing bulk contamination of $^{210}$Pb---originating from the $^{238}$U decay chain---present in archaeological Pb ingots may be further concentrated during purification, thereby degrading the overall radiopurity of the material.

For these reasons, a fast and sensitive radiopurity monitoring technique represents a valuable tool during the refining and processing of archaeological Pb intended for use in rare--event physics experiments.

The Pb employed for this study is retrieved from the very same source of the Pb employed for the construction of the CUORE cryogenic Pb shielding~\cite{ALDUINO20199}, and it is the same material that is currently being used for the production of the PbWO$_4$ crystals for the RES-NOVA experiment~\cite{RESNOVA_Pb}.
In the next sections, the application and effectiveness of the fast detection technique developed in this work will be described.

\section{Experimental setup}

\begin{figure}[h]
    \centering
    \includegraphics[width=0.9\linewidth]{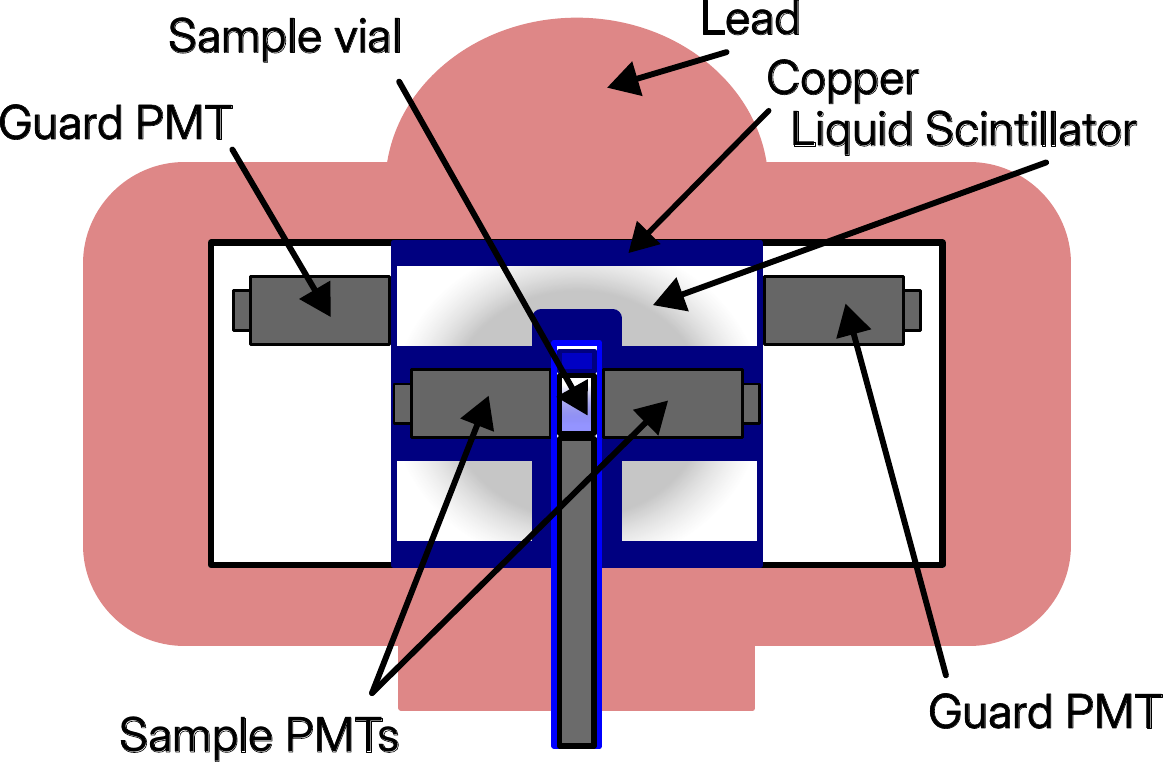}
    \caption{Scheme of the Quantulus 1220 detector where the sample site, the sample PMTs, the active muon veto system and the passive lead shielding are shown.}
    \label{schema}
\end{figure}

All measurements presented in this work were carried out using the PerkinElmer Wallac Quantulus~1220 detector installed at the low--background laboratory of the Physics Department of the University of Milano--Bicocca. This instrument is a commercial low--background liquid scintillation counter primarily employed for the measurement of $\beta$ and $\alpha$ decays from radioactive contaminants in environmental~\cite{Quantulus_sea2016rapid} and geological~\cite{Quantulus_geo2004simple} samples.

\begin{figure*}
    \centering
    \caption{Photos the climatic chamber (left) and the Quantulus 1220 detector housed inside (right).}
    \begin{subfigure}{0.48\linewidth}

        \includegraphics[width=\linewidth]{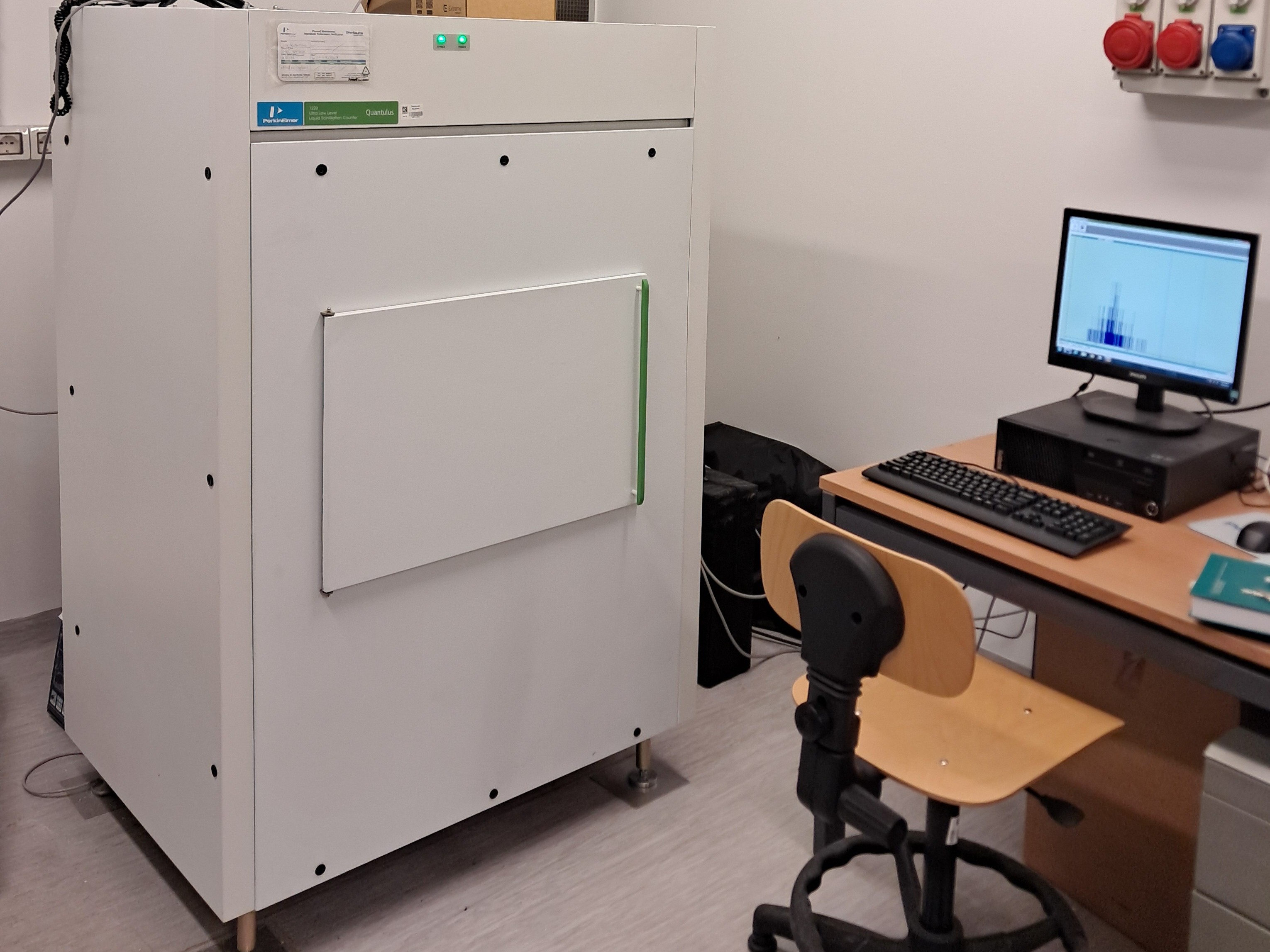}
        \caption{Photo of the climatic chamber housing the Quantulus 1220 detector and its data acquisition computer showing a live $\alpha$ spectrum of a spiked Pb sample.}
        \label{fig:frigo}
    \end{subfigure}
    \hfill
    \begin{subfigure}{0.48\linewidth}

        \includegraphics[width=\linewidth]{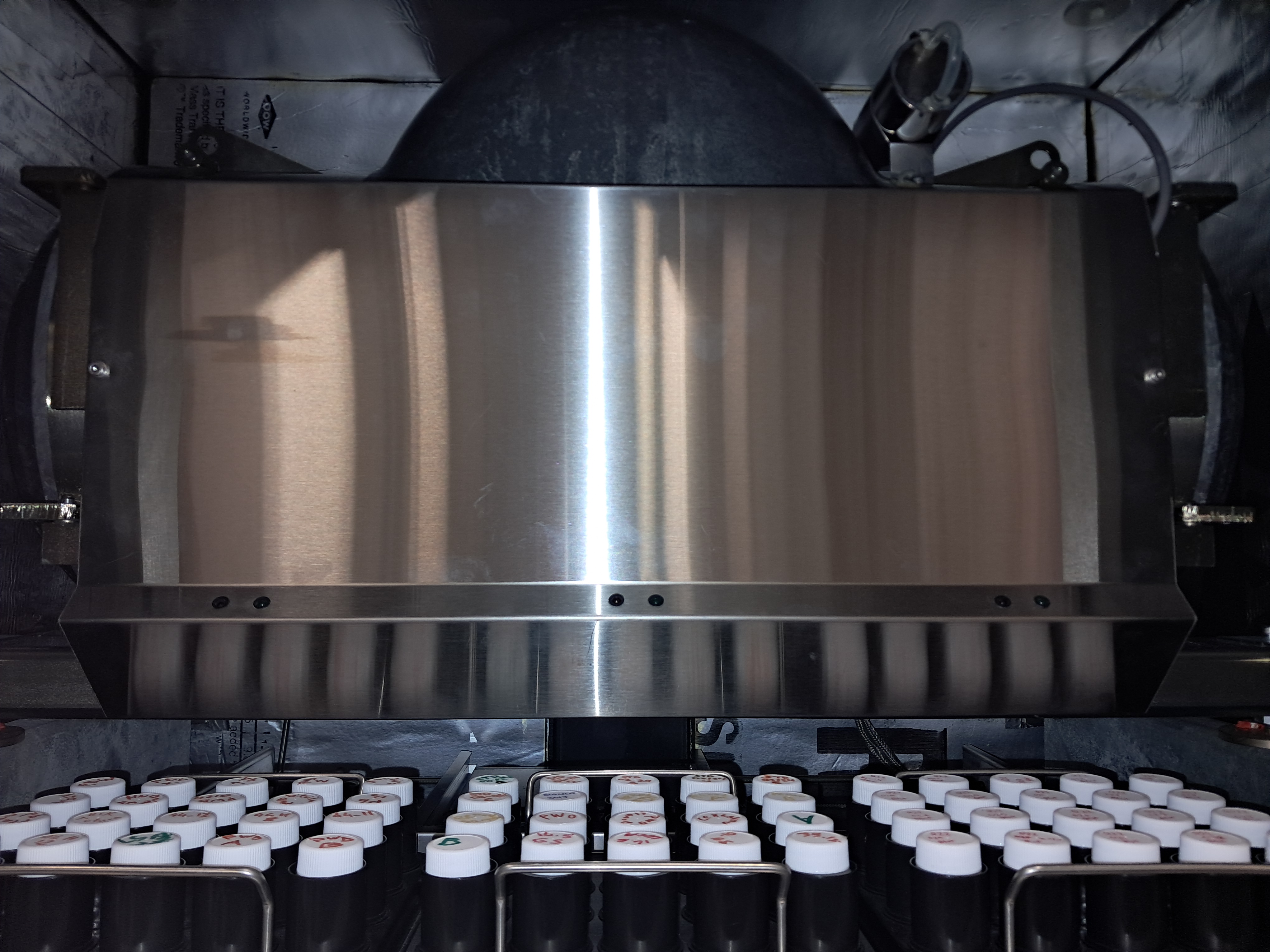}
        \caption{Photo of the Quantulus 1220 detector schematically represented in Fig. \ref{schema}. The movable sledges housing the 20 ml vials are also visible below. On the right and left sides of the metallic plate it is possible to see the lead shielding containing the coincidence PMTs.}
        \label{fig:dentro}
    \end{subfigure}
    \label{fig:foto}
\end{figure*}

The detector is designed to measure liquid samples contained in vials made of glass, quartz, or PTFE, with volumes of up to 20~ml. The vials are held by one of three movable sledges and positioned by a piston between two low--background photomultiplier tubes (PMTs), providing nearly full solid--angle coverage. The apparatus is fully enclosed in Pb shielding, except for the three movable sledges, which remain outside the shielding volume. Above the sample chamber a volume of liquid scintillator (LS), monitored by two guard PMTs, is used as an active veto against cosmic rays. The low--background PMTs are operated in coincidence to further suppress the detector background. Only signals detected simultaneously by both PMTs (e.g.\ radioactive decays occurring in the vial) are accepted, while accidental or single--PMT events are rejected. The detector and the three movable sledges are housed in a climatic chamber with a temperature of $\sim$20°C (Fig. \ref{fig:frigo}) to ensure optimal preservation of the samples. A schematic view of the detector is shown in Fig.~\ref{schema}. A photo of the detector is shown in Fig.~\ref{fig:dentro}.

The Quantulus~1220 records both accepted and rejected events. Accepted events correspond to signals detected simultaneously by the two PMTs, whereas rejected events are those detected by only one PMT. In addition, the system enables discrimination between $\alpha$ and $\beta$ particles by analyzing the decay time of the accepted scintillation signals. Due to quenching effects in the LS, $\alpha$ particles produce a light yield equivalent to that of $\beta$ particles with approximately one--tenth of the deposited energy. However, a larger fraction of the scintillation light induced by $\alpha$ particles is emitted through the slow component of the scintillation decay. This difference provides the basis for $\alpha/\beta$ discrimination via pulse--shape analysis (PSA)~\cite{knoll2010radiation}.

Since energy quenching in the LS depends on the chemical conditions of the sample (e.g.\ acidity, LS--to--digested--sample volume ratio, concentration of the dissolved material and temperature), energy calibration of the Quantulus~1220 requires a dedicated and systematic study. In practice, this involves the preparation of calibration samples with chemical conditions as close as possible to those of the samples of interest, in which well--known $\alpha$ and $\beta$ emitters are dissolved.

An additional characteristic of the Quantulus~1220 is its logarithmic analog--to--digital conversion, which provides an approximately constant relative energy resolution over a wide dynamic range but further complicates the energy calibration procedure. For these reasons, all spectra presented in this work are reported in arbitrary units.

Samples are prepared using different ratios of dissolved sample and LS, most commonly 15~ml of LS mixed with 5~ml of liquid sample  (referred to as the 5/20 sample set in this work), which is the standard ratio mainly used for aqueous matrices \cite{piraner2023alpha}, or 12~ml of LS mixed with 8~ml of liquid sample (the 8/20 sample set), which is the sample-to-LS ratio used to maximize detection sensitivity \cite{martinez2025validation}. To identify the experimental conditions that maximize sensitivity, a series of preliminary measurements was performed on both the 5/20 and 8/20 sample sets under different chemical conditions. 

All preliminary measurements were carried out without $\alpha/\beta$ discrimination. From these studies, we found that the configuration providing the highest experimental sensitivity corresponds to the 8/20 sample set.

To maximize the experimental sensitivity, it is necessary to maximize the product $m \cdot \epsilon$, where $m$ is the sample mass and $\epsilon$ is the detection efficiency. Since approximately 0.1~g of archaeological Pb can be dissolved in 1~ml of 4~M HNO$_3$, the resulting sample masses are $m_{5/20} = 0.5~\mathrm{g}$ and $m_{8/20} = 0.8~\mathrm{g}$. The quenching effect is primarily governed by the sample acidity. Assuming that the solution added to the LS has an HNO$_3$ concentration of approximately 15\%, the detection efficiencies inferred from Fig.~\ref{HNO3_conc} yield $m_{5/20} \cdot \epsilon_{5/20} = 0.40$ and $m_{8/20} \cdot \epsilon_{8/20} = 0.57$. Further details are discussed in the following sections.

Once the optimal experimental configuration was identified, a series of dedicated measurements was performed on the best achievable samples, i.e.\ those with the highest possible mass of dissolved archaeological Pb, in order to determine the optimal pulse--shape analysis (PSA) settings. A calibration campaign was then carried out on the same samples using the optimized PSA acquisition parameters. Finally, measurements on different low--background Pb samples were performed.

\section{Optimization of the experimental parameters}

The first step required to maximize the experimental sensitivity of the setup is the determination of the optimal diluted sample mass (expressed in Bq/kg), which minimizes the decision limit (DL), defined as follows~\cite{Currie1968}:

\begin{equation} 
    \mathrm{DL} \left[ \mathrm{\frac{Bq}{kg}} \right] = \frac{k_{\alpha}\, \sigma_s}{m\, \epsilon\, T\, \mathcal{B}} \; ,
    \label{sensitivity_eq}
\end{equation}

where $m$ is the sample mass in kg, $\epsilon$ is the detection counting efficiency, $T$ is the measurement time, $\mathcal{B}$ is the branching ratio of the observed decay channel, and $\sigma_s$ is the statistical fluctuation of the background in the region of interest (ROI) under the hypothesis that no signal is  present. The parameter $k_{\alpha}$ is a figure of merit that defines the confidence level of the claim in the absence of a signal ($k_{\alpha}$=2.365). In this work, all reported DL values correspond to a confidence level of 95\% one-sided.

The full optimization procedure involves the following steps, which are described in more detail in the subsequent sections:

\begin{enumerate}
    \item Selection of the optimal ratio between liquid sample and LS for the preparation of the scintillation cocktail. This is achieved by measuring the total counting efficiency of samples prepared with different LS volumes, solvent acidities, and archaeological Pb concentrations. This step aims at maximizing the product $m \cdot \epsilon$ and verifying that the chosen chemical conditions---in particular acidity and Pb concentration---allow the preparation of a homogeneous sample--LS mixture.
    
    \item Once the optimal configuration is identified, pulse--shape analysis (PSA) must be enabled, since both $\alpha$ and $\beta$ particles are detected. To determine the optimal PSA working parameters, two optimized samples are prepared and spiked with pure $\alpha$ and $\beta$ emitters, allowing the particle identification performance to be evaluated for different PSA settings.
    
    \item After identifying the optimal PSA parameters, new calibration samples are prepared under the same optimal conditions to measure the detection efficiencies for $^{210}$Pb, $^{210}$Bi, and $^{210}$Po with PSA enabled, and to define the corresponding ROIs.
    
    \item Finally, after completing the PSA efficiency calibration, the archaeological Pb samples are measured.
\end{enumerate}

\subsection{Sample preparation}

For sample preparation, a scintillation cocktail composed of dissolved Pb sample and Ultima Gold AB liquid scintillator (LS, Revvity Health Sciences Inc.) was prepared.  
The following procedure was adopted for the preparation of archaeological Pb samples:

\begin{enumerate}
    \item Approximately 2--3~g of Pb splinters are removed from re-cast archaeological Pb ingots.
    
    \item For each gram of Pb, a total solution volume of 10~ml is prepared using 3~ml of ultrapure 4~M HNO$_3$ (obtained by dilution of high--purity 65\% HNO$_3$) and 7~ml of Milli--Q ultrapure water.
    
    \item The Pb samples are dissolved in the acidic solution using a heated ultrasonic bath operated at 50~$^\circ$C.
    
    \item After complete dissolution, aliquots of Milli--Q water or ultrapure HNO$_3$ are added to the liquid sample to decrease or increase the acidity and to adjust the Pb concentration.
    
    \item Aliquots of 5~ml or 8~ml of the dissolved solution are then mixed with 15~ml or 12~ml of LS, respectively, in 20~ml PTFE vials. The final cocktail volume is 20~ml, corresponding to the 5/20 and 8/20 sample configurations.
    
    \item Each vial used for sensitivity and efficiency calibration is spiked with a certified standard solution of $^{210}$Pb.
\end{enumerate}

For both the 5/20 and 8/20 configurations, three different subsets of samples were prepared:

\begin{itemize}
    \item \textbf{s1}: samples containing different masses of dissolved archaeological Pb, spiked with 7~Bq of $^{210}$Pb tracer in a 4\% HNO$_3$ solution.
    
    \item \textbf{s2}: samples prepared with different HNO$_3$ concentrations, spiked with a fixed activity of $^{210}$Pb tracer (7~Bq) and containing a constant archaeological Pb mass of 0.05~g.
    
    \item \textbf{s3}: samples prepared with different spiked $^{210}$Pb activities at fixed HNO$_3$ concentration and archaeological Pb mass (see Table~\ref{table:S3}).
\end{itemize}

\begin{table*}[t]
\centering
\caption{Sample set S3: different $^{210}$Pb spike activities for both 5/20 and 8/20 sample configurations.}
\label{table:S3}
\begin{tabular}{lcccc}
\hline
Sample & LS [ml] & Pb mass [g] & HNO$_3$ [\%] & $^{210}$Pb activity [Bq] \\
\hline
S3\_1  & 15 & 0.039 & 3.90 & $53 \pm 4$ \\
S3\_2  & 15 & 0.039 & 3.90 & $2.7 \pm 0.2$ \\
S3\_3  & 15 & 0.039 & 3.90 & $0.68 \pm 0.06$ \\
S3\_4  & 15 & 0.039 & 3.90 & $0.17 \pm 0.02$ \\
S3\_5  & 15 & 0.039 & 3.90 & $0.034 \pm 0.004$ \\
S3\_6  & 15 & 0.039 & 3.90 & $0.012 \pm 0.001$ \\
S3\_7  & 12 & 0.063 & 2.74 & $53 \pm 4$ \\
S3\_8  & 12 & 0.063 & 2.74 & $2.7 \pm 0.2$ \\
S3\_9  & 12 & 0.063 & 2.74 & $0.68 \pm 0.06$ \\
S3\_10 & 12 & 0.063 & 2.74 & $0.17 \pm 0.02$ \\
S3\_11 & 12 & 0.063 & 2.74 & $0.034 \pm 0.004$ \\
S3\_12 & 12 & 0.063 & 2.74 & $0.012 \pm 0.001$ \\
\hline
\end{tabular}
\end{table*}

\subsection{Total efficiency measurements}

\begin{figure}[t]
    \centering
    \includegraphics[width=1.0\linewidth]{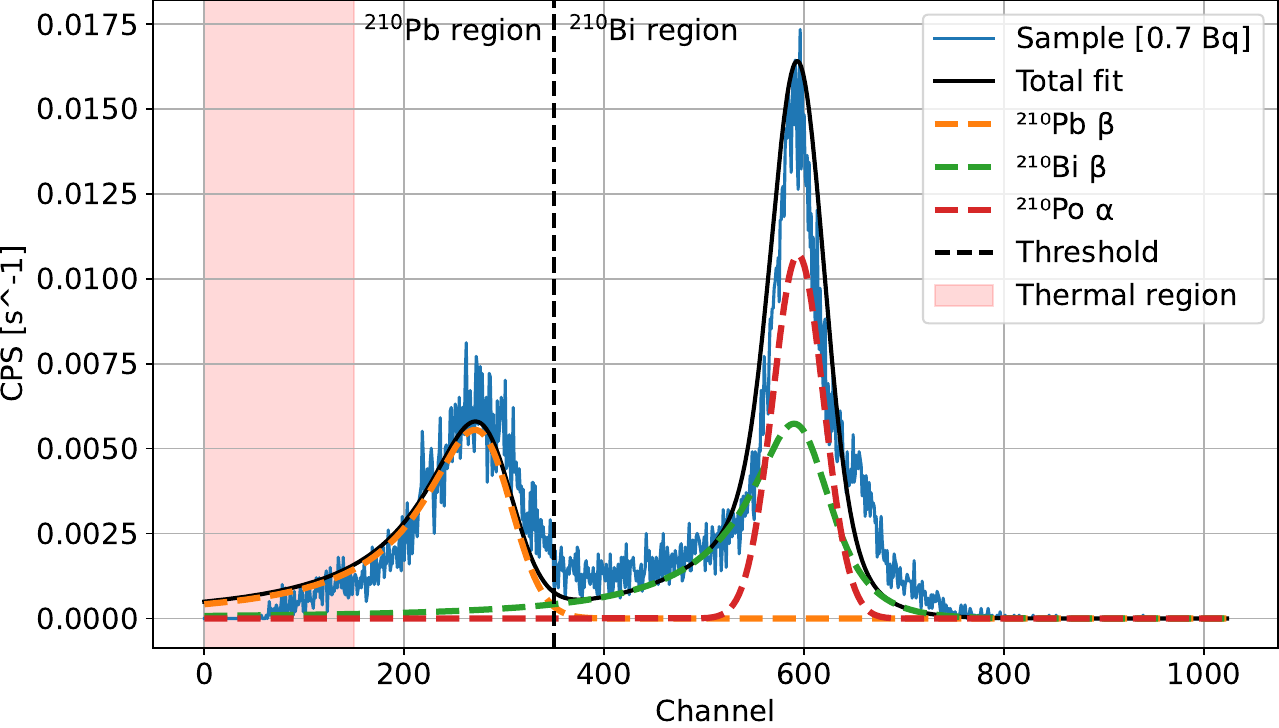}
    \caption{Detector energy spectrum acquired for the S3\_9 calibration sample. 
    The overlaid dashed curves are purely qualitative and are intended only to guide the eye. 
    The orange and green curves are asymmetric Lorentzian functions representing the contributions from the $^{210}$Pb and $^{210}$Bi $\beta$ decays, respectively.
    The red curve is a Gaussian function corresponding to the $^{210}$Po $\alpha$ contribution, which appears superimposed on the $^{210}$Bi $\beta$ continuum due to quenching effects. 
    The black solid curve represents the reconstructed total spectrum obtained as the sum of the individual components. This decomposition is not unique and is shown only to illustrate the strong spectral overlap of the different contributions.
    The reconstructed spectrum does not perfectly reproduce the data due to the absence of a detector background model. 
    An excess of events at very low energies (highlighted by the transparent red "thermal region") is also visible and is attributed to detector thermal noise.
    The vertical dashed black line indicates the chosen channel threshold separating the $^{210}$Pb and $^{210}$Bi regions.}
    \label{noPSA-description}
\end{figure}

For both the 5/20 and 8/20 initial sample subsets, the Quantulus~1220 was first operated as a pure radiation counter, without discrimination between $\alpha$ and $\beta$ particles. In this configuration, only the total number of detected events, $N_{\mathrm{counts}}$, was considered. A representative energy spectrum acquired from a spiked sample is shown in Fig.~\ref{noPSA-description}.  
This preliminary set of measurements was aimed at assessing the impact of different chemical conditions on quenching effects and on the overall detection efficiency.

\begin{figure*}[h]
    \centering
    \caption{Effects of different quenching parameters on the detector counting efficiency. In both panels, the 8/20 (5/20) sample set is shown in blue (red). Each sample was spiked with a $^{210}$Pb tracer of known activity $5.3 \pm 0.5$~Bq. Error bars include both the statistical uncertainty of the measurement and the systematic uncertainty associated with the $^{210}$Pb standard.}
    \begin{subfigure}[]{0.48\textwidth}
        \centering
        \includegraphics[width=1.0\linewidth]{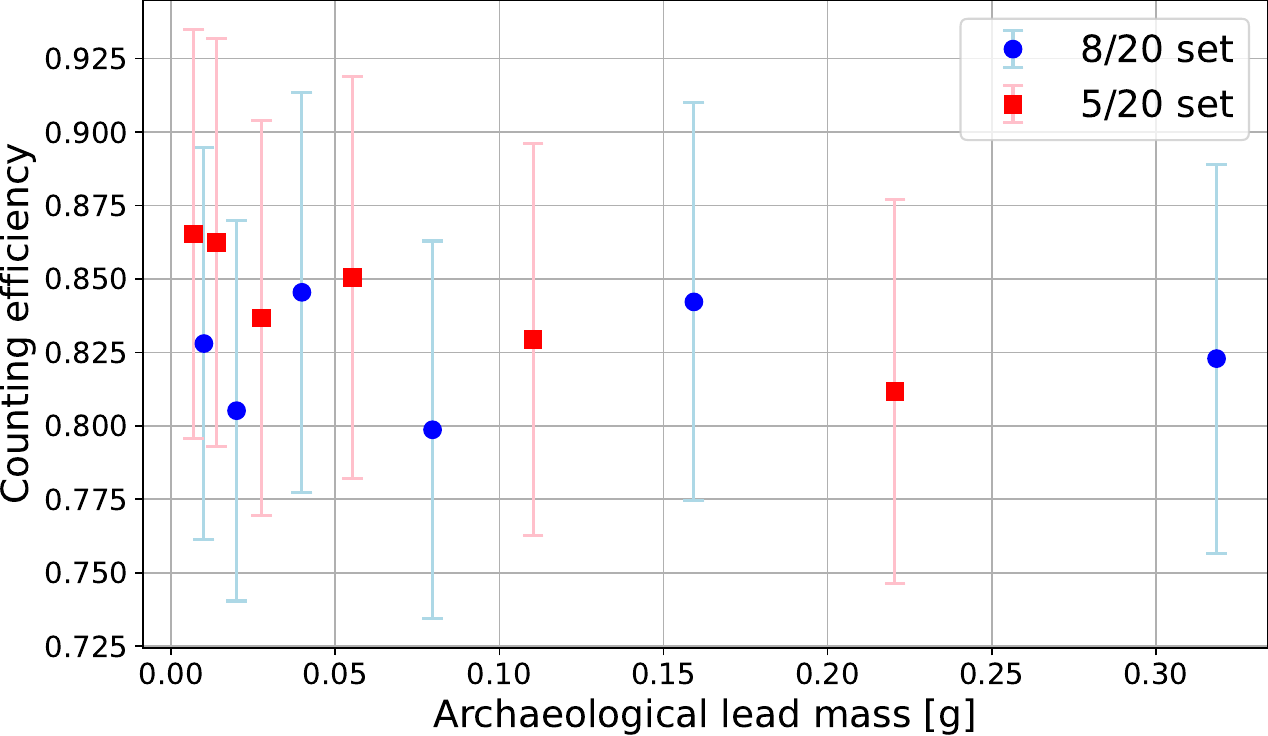}
        \caption{Detector counting efficiency as a function of the mass of digested archaeological Pb for the two sample sets. The measured efficiencies are compatible within uncertainties ($\epsilon_{5/20} = 0.84 \pm 0.02$ and $\epsilon_{8/20} = 0.82 \pm 0.02$), where the quoted uncertainties correspond to the standard deviation.}
        \label{archaeolead_masses}
    \end{subfigure}
    \hfill
    \begin{subfigure}[]{0.48\textwidth}
        \centering
        \includegraphics[width=1.0\linewidth]{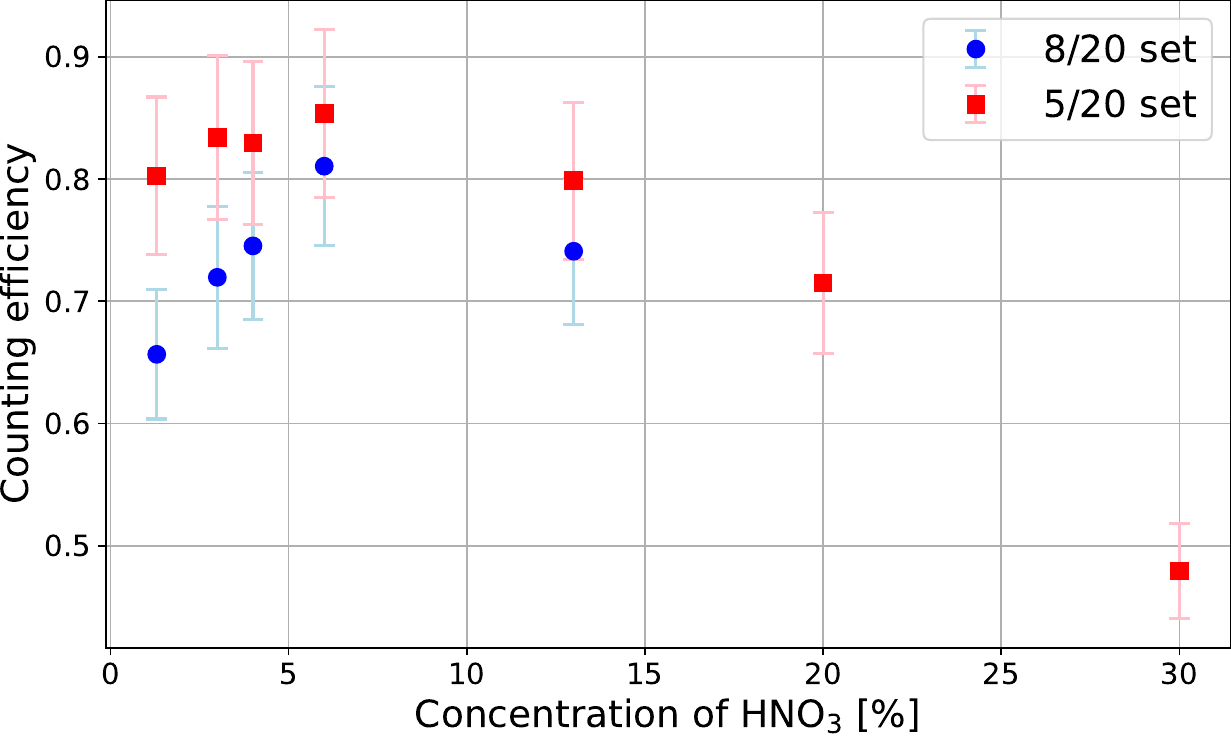}
        \caption{Effect of different HNO$_3$ concentrations in the samples on the detector counting efficiency. The 5/20 configuration exhibits a higher efficiency than the 8/20 configuration, and the efficiency response is non--constant. For both sample sets, the maximum efficiency is observed at an HNO$_3$ concentration of approximately 6\%. At higher concentrations, the efficiency decreases. No samples for the 8/20 configuration were prepared above an HNO$_3$ concentration of 13\% due to the high risk of violent exothermic reactions. Representative samples from this set are shown in Fig.~\ref{fig:fotocampioni}.}
        \label{HNO3_conc}
    \end{subfigure}
    \label{quenching}
\end{figure*}

\begin{figure}[t]
    \centering
    \includegraphics[width=0.9\linewidth]{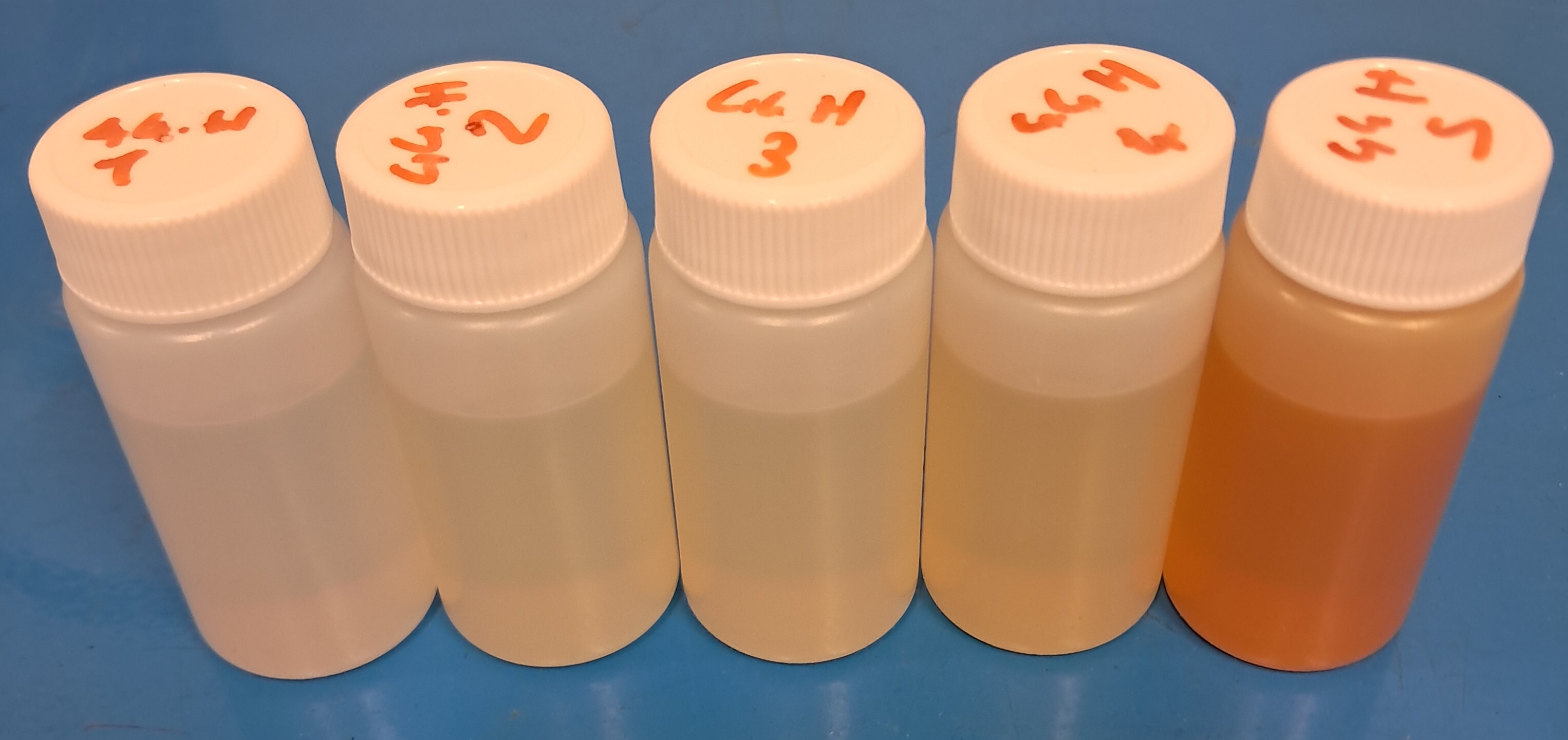}
    \caption{Photograph of the s2 sample set. From left to right, the HNO$_3$ concentration in the solution increases (from 1\% up to 13\%), and a progressive deterioration of the samples is observed. The degradation of the most acidic samples becomes visible after approximately 3--4 months from preparation, appearing initially as a slight yellow hue that darkens over time. This behavior indicates that samples prepared with the highest acidity should be discarded after a few months.}
    \label{fig:fotocampioni}
\end{figure}

The results of the efficiency ($\epsilon$) measurements are shown in Fig.~\ref{quenching}. From these plots, a mild dependence of quenching on the mass of dissolved archaeological Pb can be observed (Fig.~\ref{archaeolead_masses}), while a significantly stronger dependence on the HNO$_3$ concentration in the samples is found (Fig.~\ref{HNO3_conc}). The latter effect is also qualitatively visible through the darker coloration of LS cocktails prepared with higher acidity, as shown in figure \ref{fig:fotocampioni}.

\subsection{Optimal setup}

The aim of the preliminary study was to quantify the impact of different archaeological Pb concentrations, sample acidity, and LS cocktail volume ratios on the counting efficiency $\epsilon$. The ultimate goal is to maximize the detection sensitivity, as defined in Eq.~\ref{sensitivity_eq}.

For a fixed measurement time $T$, the sensitivity is maximized by maximizing the product $m \cdot \epsilon$. Since the efficiency response to the archaeological Pb concentration is approximately constant for both the 5/20 and 8/20 configurations, and the sample acidity is constrained by the requirement of forming a chemically stable LS cocktail, the adopted strategy to maximize sensitivity is to dissolve the largest feasible amount of archaeological Pb in the LS cocktail.
Although the 5/20 efficiency $\epsilon_{5/20}$ is higher than the 8/20 efficiency $\epsilon_{8/20}$ (as shown in Fig.~\ref{quenching}), the optimal setup is obtained using the 8/20 configuration, as it maximizes $m\cdot\epsilon$. Considering that 0.1~g of Pb can be dissolved in 1~ml of $20\%$ HNO$_3$ solution, a 5/20 sample contains approximately 0.5~g of Pb while an 8/20 sample contains 0.8~g. Even in a conservative scenario, where $\epsilon_{5/20}=0.81$ and $\epsilon_{8/20}=0.65$ (corresponding to the samples with the lowest HNO$_3$ concentration in Fig.~\ref{HNO3_conc}), $m_{8/20}\cdot\epsilon_{8/20}=0.520$ remains higher than $m_{5/20}\cdot\epsilon_{5/20}=0.405$. Therefore, the 8/20 configuration represents the optimal choice.

\subsection{PSA setup}

The Quantulus~1220 allows $\alpha/\beta$ discrimination when properly configured. Due to quenching effects, particle--dependent signal shapes are modified by the chemical conditions of the sample. As a consequence, the correct pulse--shape analysis (PSA) parameter must be determined in order to achieve an efficient separation between $\alpha$ and $\beta$ events.

Pulse Shape Analysis (PSA) is performed by integrating the tail of each scintillation pulse over a fixed time window to discriminate between short and long pulses. The resulting pulse-shape parameter is normalized to the pulse amplitude, ensuring amplitude-inde\-pendent discrimination. Events are then classified as $\alpha$- or $\beta$-like by applying a user-defined threshold in the pulse-amplitude versus pulse-shape plane, with events above (below) the threshold assigned to the long (short) pulse component \cite{manual2002wallac}. This threshold defines the PSA parameter, an integer value expressed in arbitrary units ranging from 1 to 255.
Its optimal value depends stron\-gly on the sample chemical conditions. Lower PSA values favor the identification of $\alpha$ particles but lead to an increased misidentification of $\beta$ events as $\alpha$, whereas higher PSA values improve $\beta$ identification at the expense of increased $\alpha$ misclassification.

Each prepared sample therefore has an optimal PSA value that minimizes the combined misidentification pro\-bability.

Since the Quantulus 1220 is a commercial detector, it does not provide access to the pulse shape of individual accepted events. Therefore, an \textit{a posteriori} PSA analysis cannot be performed. When the PSA function is enabled, the detector classifies each event according to the selected PSA setting and directly stores it in either the $\alpha$ or $\beta$ spectrum, which constitute the only available output data. As a consequence, the optimal PSA value for a given sample can be determined experimentally by repeating measurements of calibration samples spiked with pure $\alpha$ and $\beta$ emitters at different PSA settings \cite{piraner2023alpha}.
To determine the optimal PSA setting for the saturated 8/20 samples, three samples were prepared starting from the same mother solution of digested archaeological Pb. One sample (the $\alpha$ sample) was spiked with a pure $\alpha$ emitter ($^{238}$Pu, with an activity of $0.06 \pm 0.02$~Bq), a second sample (the $\beta$ sample) was spiked with a pure $\beta$ emitter ($^{90}$Sr, with an activity of $0.03 \pm 0.01$~Bq), and the third sample was reserved for the actual measurements with PSA enabled. The preparation of the $\alpha$ and $\beta$ calibration samples was carried out at the LENA laboratories in Pavia (Italy).

\begin{figure}[]
    \centering
    \includegraphics[width=0.9\linewidth]{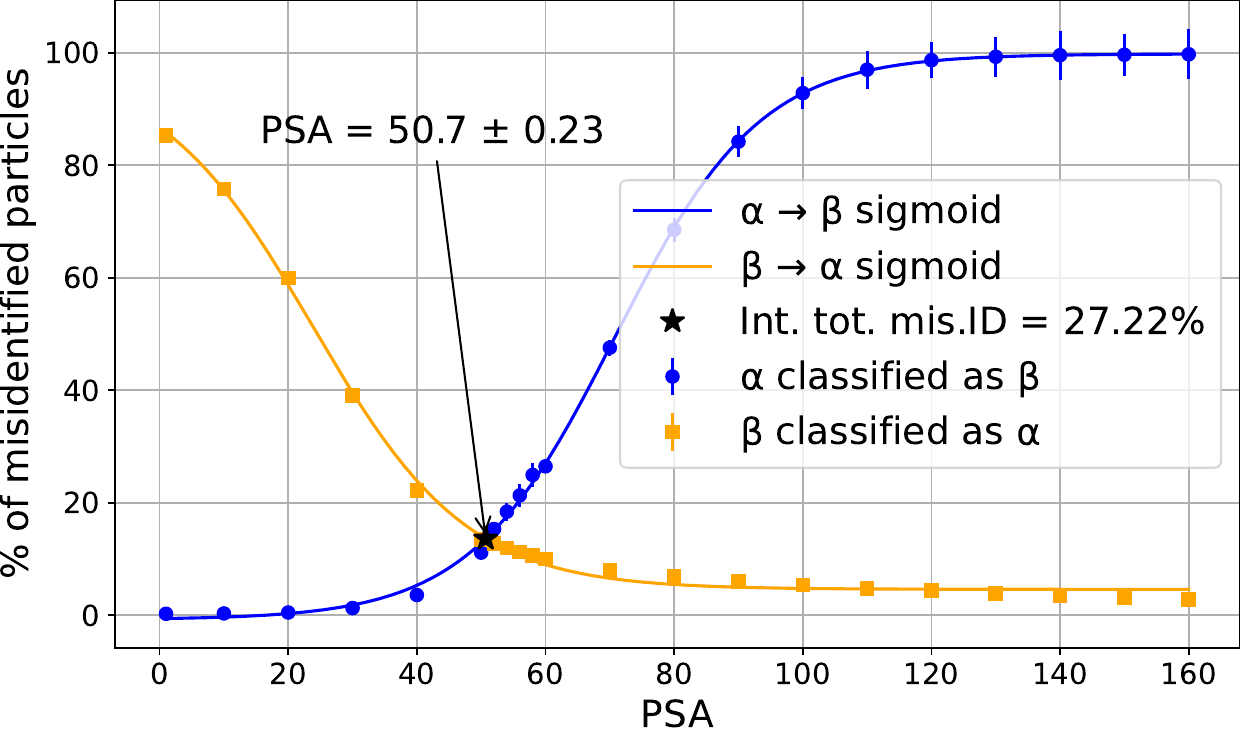}
    \caption{Misidentification probability as a function of the PSA parameter. The fraction of $\alpha$ ($\beta$) particles misidentified as $\beta$ ($\alpha$) is shown in blue (orange). The data are fitted with two sigmoid functions blue and orange. To balance $\alpha$ and $\beta$ misidentification, the optimal PSA was chosen at the intersection of the two misidentification curves. This results in a nearly symmetrical misidentification fraction (black star), even if the total number of misidentified particles is not strictly minimal. 
    Since the PSA parameter is discrete, a value of PSA~=~51 was selected.}
    \label{PSA}
\end{figure}

The $\alpha$ and $\beta$ samples were repeatedly measured using different PSA parameter values. For each measurement, the fraction of misidentified particles was evaluated. 
The optimal PSA value is, in principle, obtained by minimizing the total fraction of misidentified events. However, this condition may introduce slight asymmetries in $\alpha$ and $\beta$ identification. Since the most stringent limit is derived from the measurement of the $^{210}$Po $\alpha$ decay, a more conservative criterion that reduces the $\beta$ leakage into the $\alpha$ spectrum has been adopted. For this reason, the intersection point of the two misidentification curves is adopted as the optimal PSA value for the sample, as shown in Fig.~\ref{PSA}.

\subsection{Calibration curve with PSA}

\begin{figure*}[t]
    \centering
    \caption{Normalized $\beta$ (left) and $\alpha$ (right) spectra of saturated Pb samples spiked with different activities of a $^{210}$Pb tracer, together with a non--spiked saturated Pb sample (Archaeolead~2, discussed in the Results section) shown for reference.}
    \begin{subfigure}[t]{0.48\textwidth}
        \centering
        \includegraphics[width=1\linewidth]{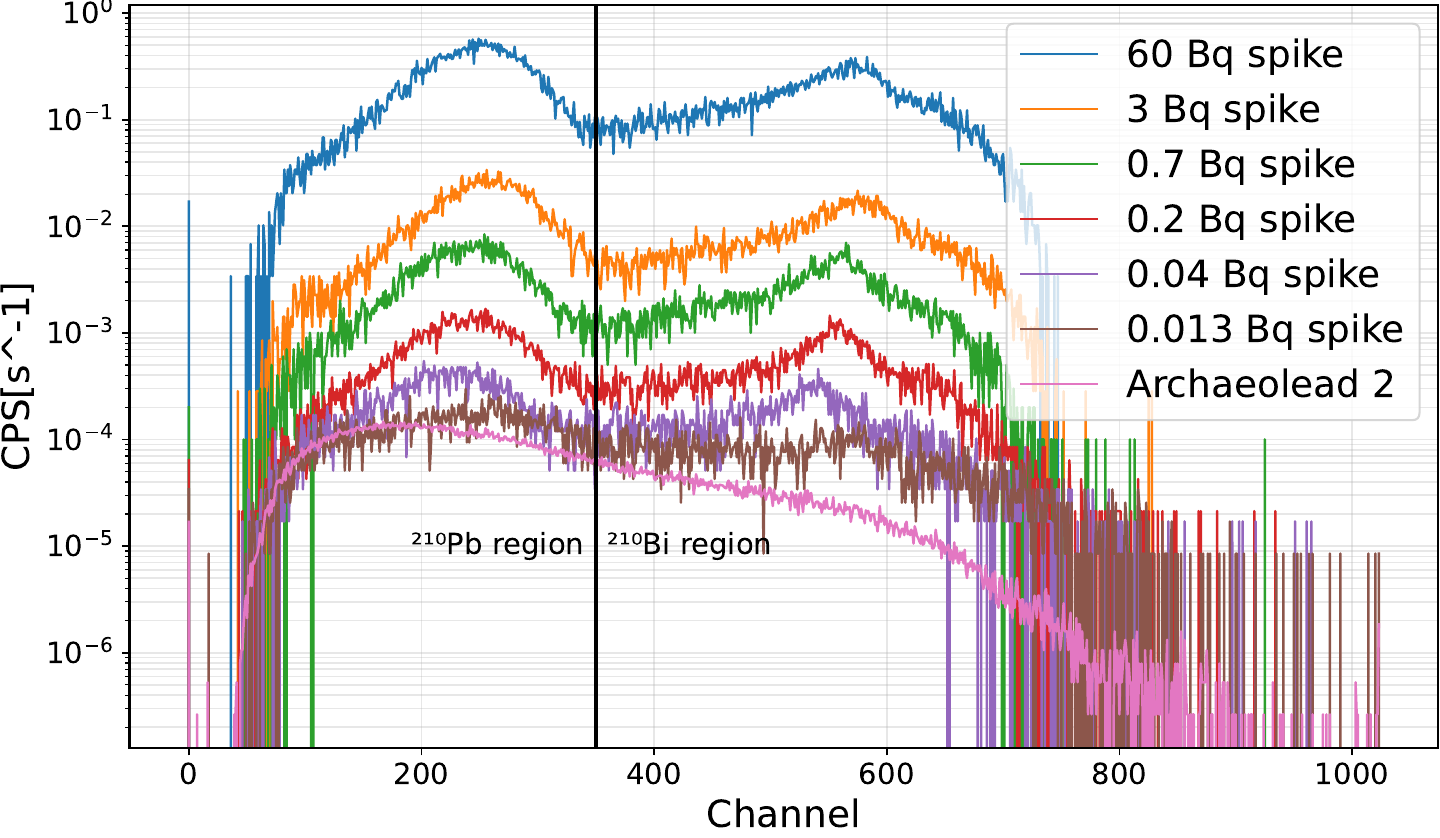}
        \caption{Definition of the $\beta$ ROIs. The threshold separating the $^{210}$Pb (left) and $^{210}$Bi (right) $\beta$ regions is set at channel~350, as indicated by the vertical black line.}
        \label{PSA_CAL_beta}
    \end{subfigure}
    \hfill
    \begin{subfigure}[t]{0.48\textwidth}
        \centering
        \includegraphics[width=1\linewidth]{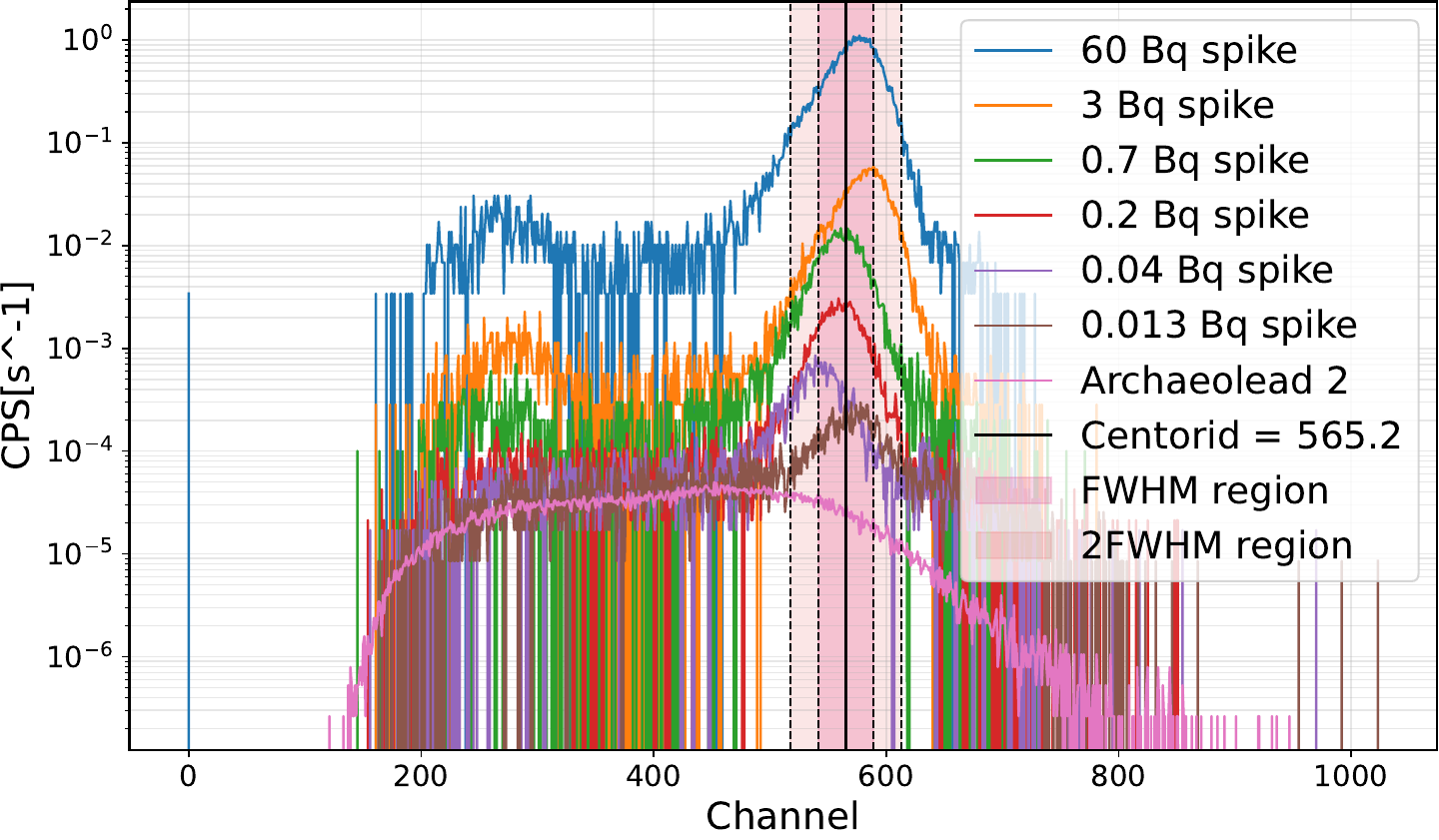}
        \caption{Definition of the $\alpha$ ROI. The 1 and 2~FWHM regions for $\alpha$ events are indicated by the vertical dotted black lines and by the darker and lighter red shaded areas, respectively. The centroid of the distribution is located at channel~562.}
        \label{PSA_CAL_alpha}
    \end{subfigure}
    \label{PSA_CAL}
\end{figure*}

Once the optimal PSA value was determined, a new set of optimized samples was prepared in order to construct a calibration curve with active particle identification enabled. The purpose of this calibration is to determine the detector counting efficiency for each decay channel and to define the regions of interest (ROIs) associated with the $\alpha$ and $\beta$ emissions of the $^{210}$Pb decay chain.

\begin{figure}[t]
    \centering
    \includegraphics[width=0.9\linewidth]{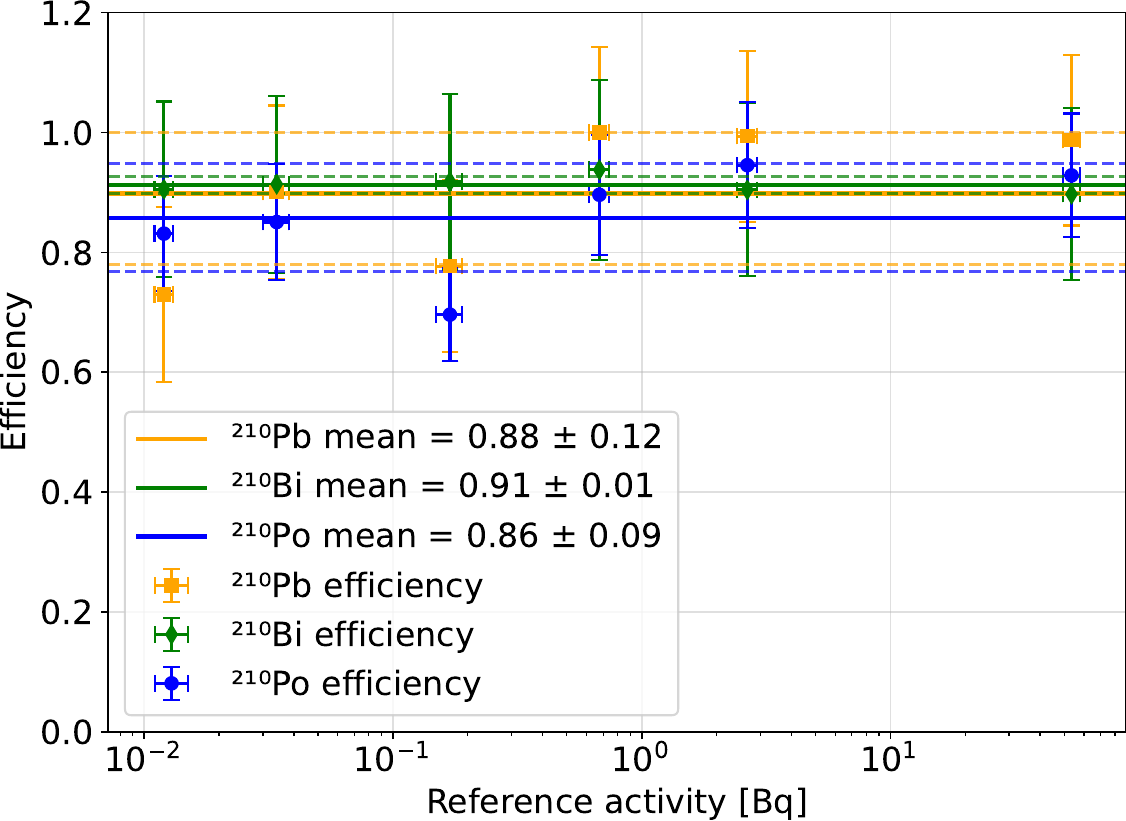}
    \caption{Counting efficiencies for each decay channel of the $^{210}$Pb decay chain, obtained for all PSA calibration samples. Efficiencies are calculated by integrating the number of events within the $\beta$ ROIs, or by integrating the fitted Gaussian peak for $\alpha$ events. All integrals are performed on background--subtracted spectra, where the background is defined by a non--spiked saturated Pb sample. The measured activities are compared with the known activities of the calibration samples. The mean counting efficiencies adopted in the activity determination for each decay channel are indicated by the horizontal dotted lines.}
    \label{fig:PSA_efficiencies}
\end{figure}

The spectra used to identify the ROIs and to extract the detection efficiencies are shown in Fig.~\ref{PSA_CAL}. The resulting counting efficiencies for the three decay channels are reported in Fig.~\ref{fig:PSA_efficiencies}.

\begin{figure}
    \centering
    \includegraphics[width=0.9\linewidth]{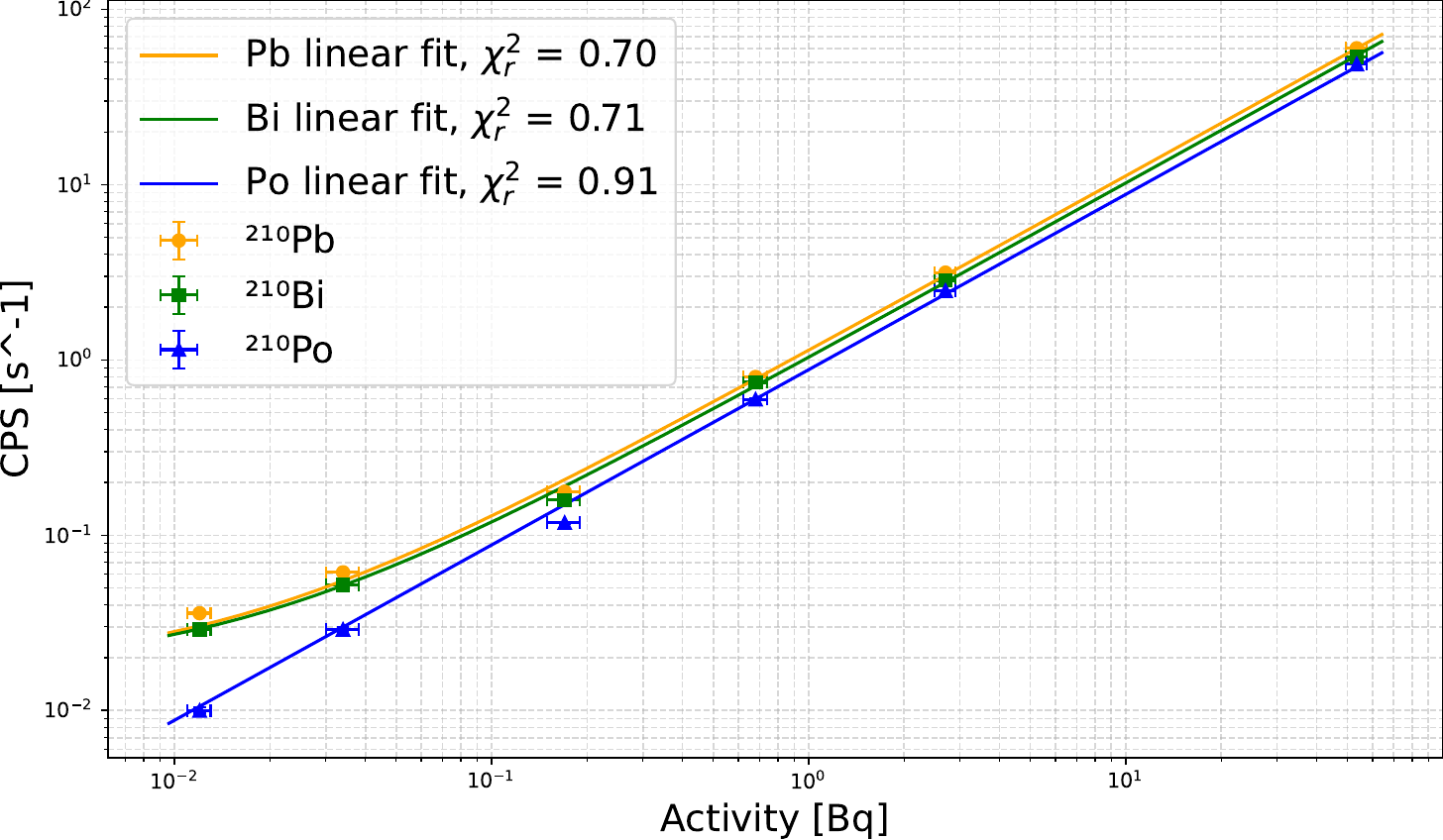}
    \caption{Counts per second (CPS) as a function of the expected activity for each isotope of the $^{210}$Pb decay chain. The $^{210}$Pb, $^{210}$Bi, and $^{210}$Po counting rates were integrated within the RoI. A logarithmic scale is used to illustrate the linear response over the full range of tested activities.
}
    \label{fig:linearity}
\end{figure}

As a final calibration step the detector linearity response was tested: all isotopes activities from the spiked samples were measured and confronted to the expected activity. The detector linearity is visible in a range of three orders of magnitude, as shown in figure \ref{fig:linearity}.
The apparent higher linearity of the $^{210}$Po calibration curve is due to the graph logarithmic scale, since the angular coefficient of the line is $< 1$ and the $^{210}$Pb  and $^{210}$Bi calibration line angular coefficients are $> 1$ ($1.12 \pm 0.05$, $1.02 \pm 0.04$ and $0.88 \pm 0.04$ for Pb, Bi and Po respectively). 

\section{Results}

\begin{figure}[]
    \centering
    \includegraphics[width=0.9\linewidth]{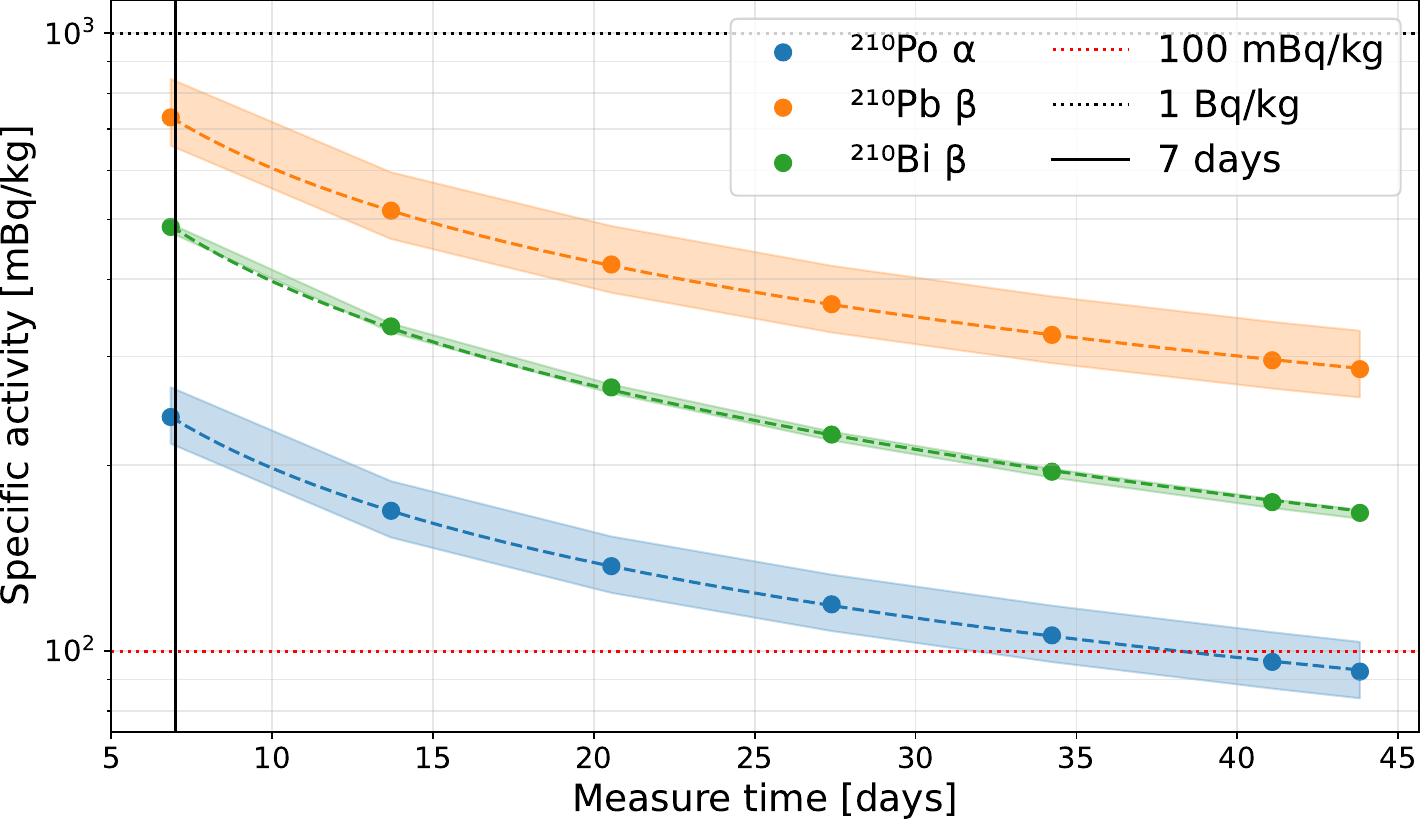}
    \caption{Evolution of the decision limit (DL) for the three decay channels of the $^{210}$Pb decay chain for the Archaeolead~2 sample. The shaded bands represent the $\pm1\sigma$ uncertainty associated with the detection efficiencies. As shown, sensitivities well below 1~Bq/kg can be achieved within one week of measurement time and a sensitivity below 100~mBq/kg for the $^{210}$Po specific activity is achieved after 40~days. Archaeolead 1 results show the same behaviour of the results shown in figure.}
    \label{fig_Archaeolead2}
\end{figure}

After completion of the full calibration procedure, four Pb samples were characterized:

\begin{itemize}
    \item \textbf{Archaeolead~1}: total archaeological Pb mass of 0.79~g. The sample was taken from a Roman Pb ingot re--cast from ancient Roman lead originally recovered from a shipwreck near Mal di Ventre (Sardinia, Italy)~\cite{clemenza2017development}. This material was previously used as shielding in the CUORE experiment and will be employed for the growth of PbWO$_4$ crystals for the RES--NOVA experiment. The sample was measured for approximately 42~days, divided into six consecutive periods of 7~days each.
    
    \item \textbf{Archaeolead~2}: total archaeological Pb mass of 0.79~g. This sample was taken from the same ingot as Archaeolead~1. The measurement was carried out for approximately 44~days following the same procedure adopted for Archaeolead~1. The evolution of the measurement sensitivity is shown in Fig.~\ref{fig_Archaeolead2}.
    
    \item \textbf{High--purity Pb}: total mass of 0.75~g obtained from high--purity Pb pellets (Low Alpha Lead) produced by CSC Pure Technologies (Russia). This sample was measured for 7~days only; therefore, no sensitivity evolution curves are reported.

    \item \textbf{Opera Pb}: total mass of 0.76 g obtained from lead sheets formerly used in the OPERA experiment. This measure was carried out for 7 days to test the procedure with samples contaminated with $^{210}$Pb.
\end{itemize}

In the first three samples, a sensitivity on the $^{210}$Po specific activity at the level of a few hundred mBq/kg was achieved within only 7~days of measurement. Assuming secular equilibrium, this sensitivity directly translates into a comparable sensitivity on the $^{210}$Pb specific activity. The best results are summarized in Table~\ref{tab:best_results}.

In all these samples, it was also possible to achieve a direct sensitivity on the $^{210}$Pb specific activity below 1~Bq/kg, although with higher values compared to those obtained through $^{210}$Po $\alpha$ detection. The most stringent limit on the $^{210}$Pb activity, below 100~mBq/kg, was achieved after 44~days of measurement for the Archaeolead~2 sample. A complete summary of the results obtained in this work is reported in Table~\ref{tab:full_results}.

In the OPERA sample, it was possible to measure all three radionuclides, although not in secular equilibrium. Po behaves differently from Pb and Bi in nitric acid solutions: surface absorption and precipitation of PoO$_2$ lower the recovery efficiency \cite{danon1956solvent}. The sample was measured shortly after its preparation, therefore no time was left to the Pb chain to achieve secular equilibrium again. The contamination of all three radionuclides was found well below 100 Bq/kg, as shown in table \ref{tab:full_results}.

In this work, we have presented an improved method for the screening of $^{210}$Pb and its progeny in metallic Pb. Importantly, the methodology is directly applicable to other materials used as sensitive components in low--background experiments (e.g.\ NaI, CaWO$_4$, Si, and Ge), providing timely feedback on manufacturing and handling procedures while requiring only minimal sample masses.

\begin{table*}[]
\centering
\caption{Best achieved sensitivities for the $^{210}$Pb specific activity, assuming secular equilibrium, after 7~days and after the full measurement campaign. All limits are reported at 95\% C.L.}
\label{tab:best_results}
\begin{tabular}{lccc}
\hline
Sample & $^{210}$Pb [mBq/kg] & Sample mass [kg] & Measurement time [d] \\
\hline
Archaeolead~1 & $<$ 203 & $7.8\times10^{-4}$ & 7  \\
Archaeolead~2 & $<$ 239 & $7.9\times10^{-4}$ & 7  \\
High--purity Pb & $<$ 258 & $7.4\times10^{-4}$ & 7  \\
Archaeolead~1 & $<$ 103 & $7.8\times10^{-4}$ & 42 \\
Archaeolead~2 & $<$ 93  & $7.9\times10^{-4}$ & 44 \\
\hline
\end{tabular}
\end{table*}

\begin{table*}[]
\centering
\caption{Results of the measurement campaign, for each radionuclide of the $^{210}$Pb decay chain after 7~days and after the full measurement campaign. Limits are reported at 95\% C.L..}
\label{tab:full_results}
\begin{tabular}{lcccc}
\hline
Sample & $^{210}$Pb [mBq/kg] & $^{210}$Bi [mBq/kg] & $^{210}$Po [mBq/kg] & Time [d] \\
\hline
Archaeolead~1 & $<$ 720 & $<$ 661 & $<$ 203 & 7  \\
              & $<$ 300 & $<$ 249 & $<$ 103 & 42 \\
Archaeolead~2 & $<$ 731 & $<$ 485 & $<$ 239 & 7  \\
              & $<$ 286 & $<$ 167 & $<$ 93  & 44 \\
High--purity Pb & $<$ 802 & $<$ 636 & $<$ 258 & 7  \\
OPERA Pb & $(67\pm7)\times 10^3$ & $(58\pm7) \times 10^3$ & $(27\pm8) \times 10^3$ & 7\\
\hline
\end{tabular}
\end{table*}

\section{Conclusions}

A fast and sensitive method for the measurement of the $^{210}$Pb activity in archaeological lead samples has been developed and fully characterized using the Wallac Quantulus~1220 liquid scintillation counter.

The technique, based on optimized chemical preparation and pulse--shape analysis (PSA), enables the simultaneous detection of $\alpha$ and $\beta$ emissions from the $^{210}$Pb decay chain, achieving counting efficiencies above 80\% for all relevant decay channels.

Detection limits below 1~Bq/kg are reached within one week of measurement, while sensitivities down to approximately 100~mBq/kg are achieved after extended acquisition times, approximately a month. These results demonstrate that the proposed setup is well suited for routine radiopurity screening of lead samples.

Overall, this method represents a valuable addition to existing radiometric assay techniques. It requires sample masses below 1~g and provides a rapid, reliable, and cost--effective solution for preliminary material screening and for monitoring purification processes in the production of radiopure Pb--based components for rare--event physics experiments. Moreover, the method has the potential to be extended to other materials relevant for low--background applications that are sensitive to, or limited by, $^{210}$Pb radioactive contamination.

\begin{acknowledgements}
This work received funding from the European Union’s Horizon Europe program through the ERC grant ERC-101087295-RES-NOVA. Additional financial support was provided under the National Recovery and Resilience Plan (NRRP) by the Italian Ministry of University and Research (MUR), funded by the European Union – NextGenerationEU, within the project “Advanced techniques for a next-generation isotopically enriched cryogenic Dark Matter experiment” - 2022L2AXP2.
We gratefully acknowledge the University of Milano-Bicocca (UNIMIB) and INFN for their continuous support.
\end{acknowledgements}

\newcommand*{\doi}[1]{\href{https://doi.org/\detokenize{#1}}{DOI: \detokenize{#1}}}

%
\bibliographystyle{spphys}       
\bibliography{template.bib}

@article{RESNOVA_Pb,
  title = {Neutrino observatory based on archaeological lead},
  author = {Pattavina, Luca and Ferreiro Iachellini, Nahuel and Tamborra, Irene},
  journal = {Phys. Rev. D},
  volume = {102},
  issue = {6},
  pages = {063001},
  numpages = {16},
  year = {2020},
  month = {Sep},
  publisher = {American Physical Society},
  doi = {10.1103/PhysRevD.102.063001}
}

@article{FerreiroIachellini:2021qgu,
    author = "Ferreiro Iachellini, N. and others",
    title = "{Operation of an Archaeological Lead PbWO$_4$ Crystal to Search for Neutrinos from Astrophysical Sources with a Transition Edge Sensor}",
    eprint = "2111.07638",
    archivePrefix = "arXiv",
    primaryClass = "physics.ins-det",
    doi = "10.1007/s10909-022-02823-8",
    journal = "J. Low Temp. Phys.",
    volume = "209",
    number = "5-6",
    pages = "872--878",
    year = "2022"
}

@article{Pattavina_2021,
doi = {10.1088/1475-7516/2021/10/064},
year = {2021},
month = {oct},
volume = {2021},
number = {10},
pages = {064},
author = {Pattavina, L. and others},
title = {{RES-NOVA} sensitivity to core-collapse and failed core-collapse supernova neutrinos},
journal = {J. Cosmol. Astropart. Phys.}
}

@article{Laubenstein:2020rbe,
    author = "Laubenstein, Matthias and Lawson, Ian",
    title = "{Low Background Radiation Detection Techniques and Mitigation of Radioactive Backgrounds}",
    doi = "10.3389/fphy.2020.577734",
    journal = "Front. in Phys.",
    volume = "8",
    pages = "506",
    year = "2020"
}

@article{ALDUINO20199,
title = {The CUORE cryostat: An infrastructure for rare event searches at millikelvin temperatures},
journal = {Cryogenics},
volume = {102},
pages = {9-21},
year = {2019},
issn = {0011-2275},
doi = {https://doi.org/10.1016/j.cryogenics.2019.06.011},
author = {C. Alduino and F. Alessandria and M. Balata and D. Biare and M. Biassoni and C. Bucci and A. Caminata and L. Canonica and L. Cappelli and G. Ceruti and A. Chiarini and N. Chott and M. Clemenza and S. Copello and A. Corsi and O. Cremonesi and A. DAddabbo and S. DellOro and L. {Di Paolo} and M.L. {Di Vacri} and A. Drobizhev and M. Faverzani and E. Ferri and M.A. Franceschi and R. Gaigher and L. Gladstone and P. Gorla and M. Guetti and L. Ioannucci and Yu.G. Kolomensky and C. Ligi and L. Marini and T. Napolitano and S. Nisi and A. Nucciotti and I. Nutini and T. ODonnell and D. Orlandi and J.L. Ouellet and C.E. Pagliarone and L. Pattavina and A. Pelosi and M. Perego and E. Previtali and B. Romualdi and A. Rotilio and C. Rusconi and D. Santone and V. Singh and M. Sisti and L. Taffarello and E. Tatananni and F. Terranova and S.L. Wagaarachchi and J. Wallig and C. Zarra},
}

@article{PMO_pattavina,
	Author = {Pattavina, L. and Nagorny, S. and Nisi, S. and Pagnanini, L. and Pessina, G. and Pirro, S. and Rusconi, C. and Sch{\"a}ffner, K. and Shlegel, V. N. and Zhdankov, V. N.},
	Doi = {10.1140/epja/s10050-020-00050-x},
	Journal = {Eur. Phys. J. A},
	Number = {2},
	Pages = {38},
	Title = {Production and characterisation of a {\$}{\$}{$\backslash$}hbox {\{}PbMoO{\}}{\_}4{\$}{\$}cryogenic detector from archaeological Pb},
	Volume = {56},
	Year = {2020}
}

@Article{PMO_Korea,
AUTHOR = {Khan, Arshad and Aryal, Pabitra and Kim, Hongjoo and Lee, Moo Hyun and Kim, Yeongduk},
TITLE = {PbMoO4 Synthesis from Ancient Lead and Its Single Crystal Growth for Neutrinoless Double Beta Decay Search},
JOURNAL = {Crystals},
VOLUME = {10},
YEAR = {2020},
NUMBER = {3},
ARTICLE-NUMBER = {150},
ISSN = {2073-4352},
DOI = {10.3390/cryst10030150}
}

@article{PWO,
      author         = "Beeman, J. W. and others",
      title          = "{New experimental limits on the alpha decays of lead
                        isotopes}",
      journal        = "Eur. Phys. J.",
      volume         = "A 49",
      year           = "2013",
      pages          = "50",
      doi            = "10.1140/epja/i2013-13050-7",
      eprint         = "1212.2422",
      archivePrefix  = "arXiv",
      primaryClass   = "nucl-ex",
      SLACcitation   = "%%CITATION = ARXIV:1212.2422;%%"
}

@article{knoll2010radiation,
  title={Radiation detection and measurement},
  author={Knoll, Glenn F},
  journal={John \& Wiley Sons Inc},
  year={2010}
}

@article{neutrinoless,
  author = "Agostini, Matteo and Benato, Giovanni and Detwiler, Jason A. and Men{\'e}ndez, Javier and Vissani, Francesco",
    title = "{Toward the discovery of matter creation with neutrinoless {\ensuremath{\beta}}{\ensuremath{\beta}} decay}",
    eprint = "2202.01787",
    archivePrefix = "arXiv",
    primaryClass = "hep-ex",
    doi = "10.1103/RevModPhys.95.025002",
    journal = "Rev. Mod. Phys.",
    volume = "95",
    number = "2",
    pages = "025002",
    year = "2023"
}

@article{darkmatter,
 doi = {10.1088/1361-6633/ac5754},
year = {2022},
month = {apr},
publisher = {IOP Publishing},
volume = {85},
number = {5},
pages = {056201},
author = {Billard, J. and others},
title = {Direct detection of dark matter {APPEC} committee report},
journal = {Rep. Prog. Phys.}
}

@article{neutrinos,
 author = "Cadeddu, M. and Dordei, F. and Giunti, C.",
    title = "{A view of coherent elastic neutrino-nucleus scattering}",
    eprint = "2307.08842",
    archivePrefix = "arXiv",
    primaryClass = "hep-ph",
    doi = "10.1209/0295-5075/ace7f0",
    journal = "EPL",
    volume = "143",
    number = "3",
    pages = "34001",
    year = "2023"
}

@article{Kamland2022,
  title = {Measurement of the double-$\beta$ decay half-life of ${}^{136}$Xe with the KamLAND-Zen experiment},
  author = {Gando, A. and Gando, Y. and Hanakago, H. and Ikeda, H. and Inoue, K. and Kato, R. and Koga, M. and Matsuda, S. and Mitsui, T. and Nakada, T. and Nakamura, K. and Obata, A. and Oki, A. and Ono, Y. and Shimizu, I. and Shirai, J. and Suzuki, A. and Takemoto, Y. and Tamae, K. and Ueshima, K. and Watanabe, H. and Xu, B. D. and Yamada, S. and Yoshida, H. and Kozlov, A. and Yoshida, S. and Banks, T. I. and Detwiler, J. A. and Freedman, S. J. and Fujikawa, B. K. and Han, K. and O'Donnell, T. and Berger, B. E. and Efremenko, Y. and Karwowski, H. J. and Markoff, D. M. and Tornow, W. and Enomoto, S. and Decowski, M. P.},
  journal = {Phys. Rev. C},
  volume = {85},
  issue = {4},
  pages = {045504},
  numpages = {6},
  year = {2012},
  month = {Apr},
  doi = {10.1103/PhysRevC.85.045504}
}

@article{kg-scale,
	Author = {Beeman, J. W. and others},
    collaboration = "{RES-NOVA}",
	Doi = {10.1140/epjc/s10052-022-10656-8},
	Id = {Beeman2022},
	Isbn = {1434-6052},
	Journal = {Eur. Phys. J. C},
	Number = {8},
	Pages = {692},
	Title = "{Radiopurity of a kg-scale $PbWO_4$ cryogenic detector produced from archaeological Pb for the RES-NOVA experiment}",
	Ty = {JOUR},
	Volume = {82},
	Year = {2022}
}

@article{juno2022juno,
    author = "Abusleme, Angel and others",
    collaboration = "JUNO",
    title = "{Sub-percent precision measurement of neutrino oscillation parameters with JUNO}",
    eprint = "2204.13249",
    archivePrefix = "arXiv",
    primaryClass = "hep-ex",
    doi = "10.1088/1674-1137/ac8bc9",
    journal = "Chin. Phys. C",
    volume = "46",
    number = "12",
    pages = "123001",
    year = "2022"
}

@article{Clemenza:2011zz,
    author = "Clemenza, M. and Maiano, C. and Pattavina, L. and Previtali, E.",
    title = "{Radon-induced surface contaminations in low background experiments}",
    doi = "10.1140/epjc/s10052-011-1805-0",
    journal = "Eur. Phys. J. C",
    volume = "71",
    pages = "1805",
    year = "2011"
}

@article{xenon2024xenonnt,
  title={The XENONnT dark matter experiment},
  author={Aprile, E and Aalbers, J and Abe, K and Ahmed Maouloud, S and Althueser, L and Andrieu, B and Angelino, E and Angevaare, JR and Antochi, VC and others},
  journal={Eur. Phys. J. C},
  volume={84},
  number={8},
  pages={784},
  year={2024},
  publisher={Springer},
  doi={10.1140/epjc/s10052-024-12982-5}
}

@article{zani2024darkside,
  title={The DarkSide-20k experiment},
  author={Zani, Andrea and DarkSide-20k Collaboration},
  journal={Journal of Instrumentation},
  volume={19},
  number={03},
  pages={C03058},
  year={2024},
  publisher={IOP Publishing}
}

@article{cuore2018first,
  title={First Results from CUORE: A Search for Lepton Number Violation via 0 $\nu$ $\beta$ $\beta$ Decay of Te 130},
  author={Alduino, Chris and Alessandria, F and Alfonso, K and Andreotti, E and Arnaboldi, C and Avignone III, FT and Azzolini, O and Balata, M and Bandac, I and Banks, TI and others},
  journal={Phys. Rev. Lett.},
  volume={120},
  number={13},
  pages={132501},
  year={2018},
  publisher={APS},
  doi={10.1103/PhysRevLett.120.132501}
}

@article{CRESSTIII2019first,
  title={First results from the CRESST-III low-mass dark matter program},
  author={Abdelhameed, Ahmed H and Angloher, G and Bauer, P and Bento, A and Bertoldo, E and Bucci, C and Canonica, L and DAddabbo, Antonio and Defay, X and Di Lorenzo, S and others},
  journal={Phys. Rev. D},
  volume={100},
  number={10},
  pages={102002},
  year={2019},
  publisher={APS},
  doi={10.1103/PhysRevD.100.102002}
}

@article{OPERA2009detection,
  title={The detection of neutrino interactions in the emulsion/lead target of the OPERA experiment},
  author={Agafonova, N and Anokhina, A and Aoki, S and Ariga, A and Ariga, T and Arrabito, L and Autiero, D and Badertscher, A and Bagulya, A and Greggio, F Bersani and others},
  journal={J. Instrum.},
  volume={4},
  number={06},
  pages={P06020},
  year={2009},
  doi={10.1088/1748-0221/4/06/P06020}
}

@article{Vaananen:2011bf,
     author         = "V{\"a}{\"a}n{\"a}nen, Daavid and Volpe, Cristina",
      title          = "{The neutrino signal at HALO: learning about the primary
                        supernova neutrino fluxes and neutrino properties}",
     journal={J. Cosmol. Astropart. Phys.},
      volume         = "1110",
      year           = "2011",
      pages          = "019",
      doi            = "10.1088/1475-7516/2011/10/019"
}

@article{pattavina2019radiopurity,
  title={Radiopurity of an archaeological Roman lead cryogenic detector},
  author={Pattavina, L and Beeman, JW and Clemenza, M and Cremonesi, O and Fiorini, E and Pagnanini, L and Pirro, S and Rusconi, C and Sch{\"a}ffner, K},
  journal={Eur. Phys. J. A},
  volume={55},
  number={8},
  pages={127},
  year={2019},
  doi={10.1140/epja/i2019-12809-0}
}

@online{lnhb_lara,
  author  = {{Laboratoire National Henri Becquerel}},
  title   = {Module LARA - DonnÃ©es nuclÃ©aires}
}

@article{alloni2025new,
  title={New dark matter direct search based on archaeological Pb},
  author={Alloni, D and Benato, G and Carniti, P and Cataldo, M and Chen, L and Clemenza, M and Consonni, M and Croci, G and Dafinei, I and Danevich, FA and others},
  journal={Phys. Rev. D},
  volume={111},
  number={10},
  pages={103050},
  year={2025},
  doi={10.1103/wcvd-rk1f}
}

@article{alessandrello1991measurements,
  title={Measurements on radioactivity of ancient roman lead to be used as shield in searches for rare events},
  author={Alessandrello, Angelo and Cattadori, Carla and Fiorentini, Giovanni and Fiorini, Ettore and Gervasio, Giovanni and Heusser, Gerd and Mezzorani, Giuseppe and Pernicka, Ernst and Quarati, Piero and Salvi, Donatella and others},
  journal={Nucl. Instrum. Methods Phys. Res. B},
  volume={61},
  number={1},
  pages={106--117},
  year={1991},
  doi={10.1016/0168-583X(91)95569-Y}
}

@article{heusser2006low,
  title={Low-level germanium gamma-ray spectrometry at the $\mu$Bq/kg level and future developments towards higher sensitivity},
  author={Heusser, Gerd and Laubenstein, M and Neder, H},
  journal={J. Environ. Radioact.},
  volume={8},
  pages={495--510},
  year={2006},
  doi={10.1016/S1569-4860(05)08039-3}
}

@article{orrell2016assay,
  title={Assay methods for 238U, 232Th, and 210Pb in lead and calibration of 210Bi bremsstrahlung emission from lead},
  author={Orrell, John L and Aalseth, Craig E and Arnquist, Isaac J and Eggemeyer, Tere A and Glasgow, Brian D and Hoppe, Eric W and Keillor, Martin E and Morley, Shannon M and Myers, Allan W and Overman, Cory T and others},
  journal={J. Radioanal. Nucl. Chem.},
  volume={309},
  number={3},
  pages={1271--1281},
  year={2016},
  doi={10.1007/s10967-016-4732-6}
}

@article{alessandrello1993measurements,
  title={Measurements of low radioactive contaminations in lead using bolometric detectors},
  author={Alessandrello, A and Allegretti, F and Brofferio, C and Camin, DV and Cremonesi, O and Fiorini, Ettore and Giuliani, A and Pavan, M and Pessina, G and Pizzini, S and others},
  journal={Nucl. Instrum. Methods Phys. Res. B},
  volume={83},
  number={4},
  pages={539--544},
  year={1993},
  doi={10.1016/0168-583X(93)95884-8}
}

@article{alessandrello1998measurements,
  title={Measurements of internal radioactive contamination in samples of Roman lead to be used in experiments on rare events},
  author={Alessandrello, A and Arpesella, C and Brofferio, C and Bucci, C and Cattadori, C and Cremonesi, O and Fiorini, Ettore and Giuliani, A and Latorre, S and Nucciotti, A and others},
  journal={Nucl. Instrum. Methods Phys. Res. B},
  volume={142},
  number={1-2},
  pages={163--172},
  year={1998},
  doi={10.1016/S0168-583X(98)00279-1}
}

@article{AMoRE-2024,
    author = "Agrawal, A. and others",
    collaboration = "AMoRE",
    title = "{Improved Limit on Neutrinoless Double Beta Decay of Mo100 from AMoRE-I}",
    eprint = "2407.05618",
    archivePrefix = "arXiv",
    primaryClass = "nucl-ex",
    doi = "10.1103/PhysRevLett.134.082501",
    journal = "Phys. Rev. Lett.",
    volume = "134",
    number = "8",
    pages = "082501",
    year = "2025"
}

@article{Boiko,
	Author = {Boiko, R. S. and Virich, V. D. and Danevich, F. A. and Dovbush, T. I. and Kovtun, G. P. and Nagornyi, S. S. and Nisi, S. and Samchuk, A. I. and Solopikhin, D. A. and Shcherban', A. P.},
	Doi = {10.1134/S0020168511060069},
	Journal = {Inorg. Mater.},
	Number = {6},
	Pages = {645--648},
	Title = {Ultrapurification of archaeological lead},
	Volume = {47},
	Year = {2011}
}

@article{Quantulus_sea2016rapid,
  title={Rapid determination of 210Pb and 210Po in water and application to marine samples},
  author={Villa-Alfageme, Maria and Mas, JL and Hurtado-Bermudez, S and Masqu{\'e}, Pere},
  journal={Talanta},
  volume={160},
  pages={28--35},
  year={2016},
  publisher={Elsevier}
}

@article{Quantulus_geo2004simple,
  title={A simple method for 210Pb determination in geological samples by liquid scintillation counting},
  author={Blanco, P and Lozano, JC and Escobar, V G{\'o}mez and Tom{\'e}, F Vera},
  journal={Applied Radiation and Isotopes},
  volume={60},
  number={1},
  pages={83--88},
  year={2004},
  publisher={Elsevier}
}

@misc{manual2002wallac,
  title={wallac 1220 Quantulus Ultra Low Level Liquid Scintillation Spectrometer},
  author={PerkinElmer},
  year={2002},
  publisher={PerkinElmer}
}

@article{clemenza2017development,
  title={Development of a multi-analytical approach for the characterization of ancient Roman lead ingots},
  author={Clemenza, Massimiliano and Contini, Alessandro and Baccolo, Giovanni and di Vacri, Maria Laura and Ferrante, Marco and Nisi, Stefano and Carpinelli, Massimo and Cremonesi, Oliviero and Enzo, Stefano and Fiorini, Ettore and others},
  journal={J. Radioanal. Nucl. Chem.},
  volume={311},
  number={2},
  pages={1495--1501},
  year={2017},
  doi={10.1007/s10967-016-5040-x}
}

@article{danon1956solvent,
  title={Solvent extraction of polonium from nitric acid solutions},
  author={Danon, J and Zamith, AAL},
  journal={Nature},
  volume={177},
  number={4512},
  pages={746--747},
  year={1956},
  publisher={Nature Publishing Group UK London}
}

@article{Currie1968,
  author = {L. A. Currie},
  title = {Limits for Qualitative Detection and Quantitative Determination},
  journal = {Analytical Chemistry},
  volume = {40},
  number = {3},
  pages = {586--593},
  year = {1968},
  doi = {10.1021/ac60259a007}
}

@article{piraner2023alpha,
  title={Alpha and beta spillover in liquid scintillation counting analysis of urine samples},
  author={Piraner, Olga and Eardley, Karlee and Button, Jonathan},
  journal={Journal of radioanalytical and nuclear chemistry},
  volume={332},
  number={9},
  pages={3837--3844},
  year={2023},
  publisher={Springer}
}

@article{martinez2025validation,
  title={Validation of a direct method for measuring 14C in water by liquid scintillation counting (according to the ISO/IEC 17025 criteria)},
  author={Mart{\'\i}nez, Joana and Trull{\`a}s, Abril and Gongora, Magdiel and Pe{\~n}alver, Alejandra and Borrull, Francesc and Aguilar, Carme},
  journal={Journal of Radioanalytical and Nuclear Chemistry},
  pages={1--12},
  year={2025},
  publisher={Springer}
}

@article{bode2021roman,
  title={Roman lead ingots from Macedonia—the Augustan shipwreck of Comacchio (prov. Ferrara, Italy) and the reinterpretation of its lead ingots’ provenance deduced from lead isotope analysis},
  author={Bode, Michael and Hanel, Norbert and Rothenh{\"o}fer, Peter},
  journal={Archaeological and Anthropological Sciences},
  volume={13},
  number={10},
  pages={163},
  year={2021},
  publisher={Springer}
}

@article{bode2009tracing,
  title={Tracing Roman lead sources using lead isotope analyses in conjunction with archaeological and epigraphic evidence—a case study from Augustan/Tiberian Germania},
  author={Bode, Michael and Hauptmann, Andreas and Mezger, Klaus},
  journal={Archaeological and anthropological sciences},
  volume={1},
  number={3},
  pages={177--194},
  year={2009},
  publisher={Springer}
}

@article{cuesta2022comparative,
  title={A comparative study of alternative methods for 210Pb determination in environmental samples},
  author={Cuesta, E and Barba-Lobo, A and Lozano, RL and San Miguel, EG and Mosqueda, F and Bol{\'\i}var, JP},
  journal={Radiation Physics and Chemistry},
  volume={191},
  pages={109840},
  year={2022},
  publisher={Elsevier}
}

@article{stojkovic2020210pb,
  title={210Pb/210bi detection in waters by cherenkov counting--perspectives and new possibilities},
  author={Stojkovi{\'c}, Ivana and Todorovi{\'c}, Nata{\v{s}}a and Nikolov, Jovana and Tenjovi{\'c}, Branislava and Gad{\v{z}}uri{\'c}, Slobodan and Tot, Aleksandar and Vrane{\v{s}}, Milan},
  journal={Radiation Physics and Chemistry},
  volume={166},
  pages={108474},
  year={2020},
  publisher={Elsevier}
}

\end{document}